\documentclass[11pt,preprint]{emulateapj}
\usepackage{graphicx,natbib}
\usepackage{url}
\usepackage{color}

\newcommand{\zphot}{$z_{\rm phot}$}
\newcommand{\zspec}{$z_{\rm spec}$}

\begin{document}   

\title{A Galaxy Photometric Redshift Catalog for the Sloan Digital Sky Survey Data Release 6}      

\author{
Hiroaki Oyaizu$^{1,2}$,
Marcos Lima$^{2,3}$,
Carlos E. Cunha$^{1,2}$,
Huan Lin$^{4}$, 
Joshua Frieman$^{1,2,4}$,
Erin S. Sheldon$^{5}$
}

\affil{
${}^{1}$Department of Astronomy and Astrophysics, University of Chicago, Chicago, IL 60637 \\
${}^{2}$Kavli Institute for Cosmological Physics, University of Chicago, Chicago, IL 60637 \\
${}^{3}$Department of Physics, University of Chicago, Chicago, IL 60637 \\
${}^{4}$Center for Particle Astrophysics, Fermi National Accelerator Laboratory, Batavia, IL 60510 \\
${}^{5}$Center for Cosmology and Particle Physics and Department of Physics, New York University, New York, NY 10003 \\
}



\begin{abstract}

We present and describe a catalog of galaxy photometric redshifts (photo-z's) 
for the  Sloan Digital Sky Survey (SDSS) Data Release 6 (DR6). 
We use the Artificial Neural Network (ANN) technique to calculate photo-z's 
and the Nearest Neighbor Error (NNE) method to estimate photo-z errors for 
$\sim$ 77 million objects classified as galaxies in DR6 with $r < 22$.  
The photo-z and photo-z error estimators are trained and validated on a 
sample of $\sim 640,000$ galaxies that have SDSS photometry and 
spectroscopic redshifts measured by SDSS, 2SLAQ, CFRS, CNOC2, TKRS, 
DEEP, and DEEP2. 
For the two best ANN methods we have tried,
we find that 68\% of the galaxies in the validation set have a photo-z
 error smaller than
$\sigma_{68} =0.021$ or $0.024$.
After presenting our results and quality tests, we provide a short guide
for users accessing the public data.

\end{abstract}



\section{Introduction}\label{int}

While spectroscopic redshifts have now been measured for over one million 
galaxies, in recent years 
digital sky surveys have obtained multi-band imaging 
for of order a hundred million galaxies. Deep, wide-area surveys planned for 
the next decade will increase the number of galaxies with 
multi-band photometry to a few billion. Due to technological and financial 
constraints, obtaining spectroscopic redshifts for more than a 
small fraction of these galaxies will remain impractical for the foreseeable 
future. As a result, over the last decade substantial effort has gone into 
developing photometric redshift (photo-z) techniques, which use 
multi-band photometry to estimate approximate galaxy redshifts. For many 
applications in extragalactic astronomy and cosmology, the resulting 
photometric redshift precision is sufficient for the science goals at 
hand, provided one can accurately characterize the uncertainties in the 
photo-z estimates. 

Two broad categories of photo-z estimators are in wide use: 
template-fitting and training set methods. In template-fitting, one 
assigns a redshift to a galaxy by finding 
the redshifted spectral energy distribution (SED), selected 
from a libary of templates, 
that best reproduces the observed fluxes in the broadband filters.
By contrast, in the training set approach, one 
uses a training set of galaxies with 
spectroscopic redshifts and photometry to derive an empirical relation
between photometric observables (e.g., magnitudes, colors, and morphological 
indicators) and redshift. 
Examples of empirical methods include Polynomial Fitting \citep{con95b}, 
the Nearest Neighbor method \citep{csa03}, 
the Nearest Neighbor Polynomial (NNP) technique \citep{cun07}, 
Artificial Neural Networks (ANN) \citep{col04,van04,dab07}, and 
Support Vector Machines \citep{wad04}. When a large spectroscopic 
training set that is representative of the photometric data set to be 
analyzed is 
available, training set techniques typically outperform template-fitting 
methods, in the sense that the photo-z estimates have smaller scatter 
and bias with respect to the true redshifts \citep{cun07}. On the 
other hand, template-fitting can be applied to a photometric sample 
for which relatively few spectroscopic analogs exist.
For a comprehensive review and comparison of photo-z methods,  
see \cite{cun07}.

In this paper, we present a publicly available galaxy photometric redshift 
catalog for the Sixth Data Release (DR6) of the Sloan Digital Sky 
Survey (SDSS) imaging catalog \citep{bla03b,eis01,gun98,ive04,str02,yor00}. 
We use the ANN photo-z method, which we have shown to 
be a superior training set method \citep{cun07}, and briefly compare the 
results using different photometric observables.
We also compare the ANN results with those from NNP, an empirical
method which achieves similar performance to the ANN method \citep{cun07}. 
Since the SDSS photometric catalog covers a large area of sky, a number 
of deep spectroscopic galaxy samples with SDSS photometry are available 
to use as training sets, as shown in Fig.~\ref{dist.sdss}. 
In combination, these spectroscopic samples cover the full apparent 
magnitude range of the SDSS photometric sample.  

The paper is organized as follows.
In \S \ref{sel} we briefly describe the SDSS DR6 photometric catalog 
and the selection criteria used
to obtain the galaxy photometric sample from the catalog. 
In \S \ref{tra} we describe the spectroscopic catalogs used 
to construct the photo-z training and validation sets. 
In \S \ref{met} we outline the photo-z methods as well as the 
photo-z error estimator technique applied to the galaxy sample. 
Statistical results for photometric redshift performance, errors, 
and redshift distributions 
are presented in \S \ref{res}. In \S \ref{rec}
we make recommendations for possible
additional cuts on the photo-z catalog based on our
own flags and those in the SDSS database.
In \S \ref{cat} we briefly describe how to access the 
photo-z catalog from the public SDSS data server, and in \S \ref{con} we  
present our conclusions. For completeness, Appendix \ref{query} 
provides the database query used to select the photometric sample, 
Appendix \ref{stargal} discusses issues of star-galaxy separation, 
and Appendix \ref{photdr5} briefly describes an earlier version 
of the photo-z algorithm used for SDSS DR5 \citep{ade07}.

\section{SDSS Photometric Catalog and Galaxy Selection}
\label{sel}

The SDSS comprises a large-area 
imaging survey of the north Galactic cap, a multi-epoch imaging survey of  
an equatorial stripe in the south Galactic cap, and a spectroscopic survey of 
roughly $10^6$ galaxies and $10^5$ quasars 
\citep{yor00}. 
The survey uses a dedicated, wide-field, 2.5m telescope \citep{gun06} at 
Apache Point Observatory, New Mexico. 
Imaging is carried out in drift-scan mode using a 142 mega-pixel camera 
\citep{gun06} that gathers data in five broad bands, $u g r i z$, spanning 
the range from 3,000 to 10,000 \AA \, \citep{fuk96}, with an effective exposure 
time of 54.1 seconds per band. 
The images are processed using specialized 
software \citep{lup01,sto02} and are 
astrometrically \citep{pie03} and photometrically \citep{hog01,tuc06} 
calibrated using observations of a set of primary standard stars 
\citep{smi02} observed on a neighboring 20-inch telescope.

The imaging in the sixth SDSS Data Release (DR6) covers an essentially 
contiguous region of the north Galactic cap, with only a few small patches 
remaining to be observed. In any region where imaging runs overlap, one run is 
declared primary\footnote{For the precise definition of primary objects see 
{\tt http://cas.sdss.org/dr6/en/help/docs/glossary.asp\#P}} 
and is used for spectroscopic target selection; 
other runs are declared secondary. 
The area covered by the DR6 primary imaging survey, including the 
southern stripes, is $8417 \textrm{ deg}^2$, but  
DR6 includes both the primary and secondary observations of 
each area and source \citep{dr6}.

\begin{figure}
  \begin{minipage}[t]{85mm}
    \begin{center}
      \resizebox{85mm}{!}{\includegraphics[angle=0]{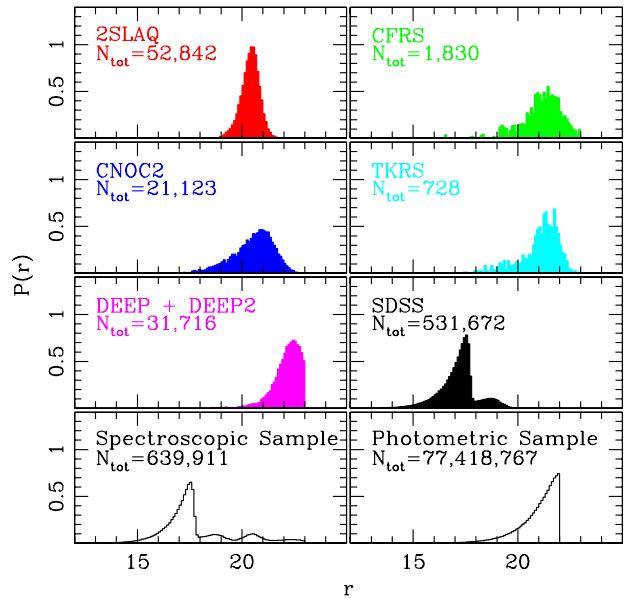}}
    \end{center}
  \end{minipage}
\caption{Normalized $r$ magnitude distributions for various catalogs.
  {\it Top three rows:} 
  the distributions of the spectroscopic catalogs used for photo-z
  training and validation are 
  shown for 2SLAQ, CFRS, CNOC2, TKRS,
  DEEP and DEEP2, and the SDSS spectroscopic sample.  
  $N_{tot}$ denotes the total number of galaxy measurements used 
  from each catalog; for galaxies in regions with repeat SDSS imaging,
  each independent photometric measurement is counted separately.
  {\it Bottom row:} ({\it left})---the distribution of the combined  
  spectroscopic sample; ({\it right})---the
  distribution for the SDSS photometric galaxy sample, where 
  objects were classified as galaxies according to the 
  photometric TYPE flag (see text).
}\label{dist.sdss}
\end{figure}


The SDSS database provides a variety of measured magnitudes for each 
detected object. Throughout this paper, we use dereddened model magnitudes to 
perform the photometric redshift computations. To determine the model 
magnitude, the SDSS photometric pipeline fits two 
models to the image of each galaxy in each passband: a de Vaucouleurs (early-type) and 
an exponential (late-type) light profile. 
The models are convolved with the estimated point 
spread function (PSF), with arbitrary axis ratio and position angle.
The best-fit model in the $r$ band (which is used to fix the model scale 
radius) is then applied to the other passbands and convolved with the 
passband-dependent PSFs to yield the model magnitudes.
Model magnitudes provide an unbiased color estimate in the absence of color 
gradients \citep{sto02}, and the dereddening procedure removes the 
effect of Galactic extinction \citep{sch98}. 


\begin{deluxetable}{c c | c c}
\tablewidth{0pt}
\tablecaption{Photometric Sample Properties}
\startdata
\hline
\hline
\multicolumn{2}{c}{\hspace{0.1 in} AB magnitude limits \hspace{0.2 in}  } 
&\multicolumn{2}{c}{\hspace{0.2 in} RMS photometric \hspace{0.4 in}} \\
\multicolumn{2}{c}{}                    
& \multicolumn{2}{c}{\hspace{0.2 in} calibration errors } \\
\hline
  \hspace{0.1 in}  $u$ & 22.0 & \hspace{0.4 in} $r$   & 2\% \\ 
  \hspace{0.1 in}  $g$ & 22.2 & \hspace{0.4 in} $u-g$ & 3\% \\ 
  \hspace{0.1 in}  $r$ & 22.2 & \hspace{0.4 in} $g-r$ & 2\% \\ 
  \hspace{0.1 in}  $i$ & 21.3 & \hspace{0.4 in} $r-i$ & 2\% \\ 
  \hspace{0.1 in}  $z$ & 20.5 & \hspace{0.4 in} $i-z$ & 3\% \\
\enddata
\tablecomments{Magnitude limits are for 95\% completeness for point
  sources in typical seeing; 50\% completeness numbers are generally
  0.4 mag fainter \citep{ade07}. The median seeing for the SDSS imaging 
  survey is $1.4''$. 
} \label{propphot}
\end{deluxetable}

To construct the photometric sample of galaxies for which we wish to 
estimate photo-z's, we obtained
a catalog drawn from the SDSS CasJobs website 
{\tt http://casjobs.sdss.org/casjobs/}.
We checked some of the SDSS photometric flags to ensure that we have obtained  
a reasonably clean galaxy sample. In particular, 
we selected all primary objects from DR6 that have the TYPE flag 
equal to $3$ (the type for galaxy) and that do not 
have any of the flags BRIGHT, SATURATED, or SATUR\_CENTER set.
For the definitions of these flags we refer the reader to the 
PHOTO flags entry at the SDSS 
website\footnote{{\tt http://cas.sdss.org/dr6/en/help/browser/browser.asp}}
or to Appendix \ref{query}.
We also took into account the nominal SDSS flux limit
(see Table~\ref{propphot}) by only selecting galaxies with dereddened model 
magnitude $r<22.0$. 
The full database query we used is given in Appendix \ref{query}.
 
The photometric galaxy catalog we have selected suffers from impurity and 
incompleteness at some level, since 
the photometric pipeline cannot 
separate stars from galaxies with 100\% success 
at faint magnitudes. We 
describe some of our tests of star/galaxy separation in 
Appendix \ref{stargal}, where we show that the SDSS TYPE flag 
provides star/galaxy separation performance similar to other
methods. 

\begin{figure}
  \begin{minipage}[t]{85mm}
    \begin{center}
      \resizebox{85mm}{!}{\includegraphics[angle=0]{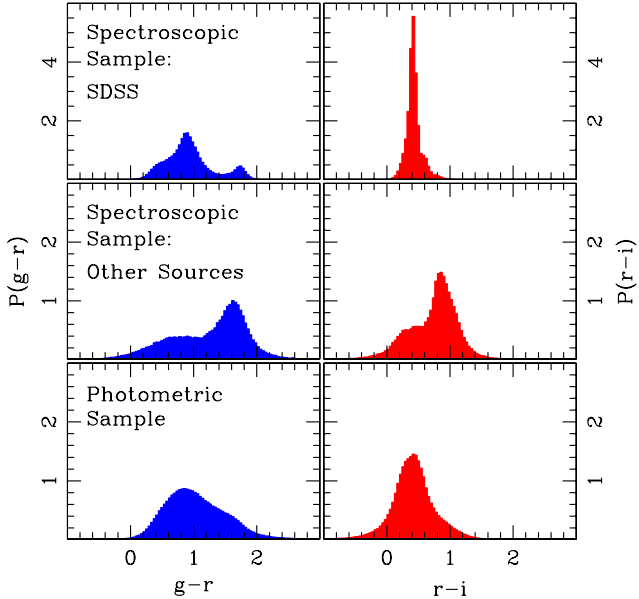}}
    \end{center}
  \end{minipage}
  \caption{Distribution of $g-r$ and $r-i$ colors for different SDSS samples. {\it Top row:} the color distributions for galaxies in the SDSS spectroscopic 
    sample. 
    {\it Middle row:} the color distributions for galaxies in the other (non-SDSS)
    spectroscopic training samples. 
    {\it Bottom row:} the color distributions for galaxies in the photometric 
    sample.
    As above, galaxy/star classification used the photometric TYPE flag.
}\label{dist.color.sdss}
\end{figure}

The final photometric sample comprises $77,418,767$ galaxies. 
The $r$ magnitude distribution of this sample is shown in 
the bottom right panel of Fig.~\ref{dist.sdss}; the $g-r$ and 
$r-i$ color distributions 
are shown in the bottom panels of Fig.~\ref{dist.color.sdss}.

\section{Spectroscopic Training and Validation sets} \label{tra}

Since our methods to estimate photo-z's and photo-z errors are 
training-set based, we would ideally like the spectroscopic 
training set to be
fully representative of the photometric sample to be analyzed, i.e., to have
similar statistical properties and magnitude/redshift distributions.  
Training-set methods can be thought of as inherently Bayesian, in the sense 
that the training-set distributions form effective priors for the analysis of the 
photometric sample; to the extent that the training-set distributions 
reflect those of the photometric sample, we may expect the photo-z estimates 
to be unbiased (or at least they will not be biased by the prior). 
Given the practical difficulties of carrying out spectroscopy at 
faint magnitudes and low surface brightness, such an ideal generally cannot be achieved.
Realistically, all we can hope for is a training set that 
(a) is large enough that statistical fluctuations are small and (b) 
spans the same magnitude, color, and redshift ranges as the photometric sample. 
Fortunately, our tests indicate that the estimated photo-z's
depend only weakly on the shape of the
redshift and magnitude distributions of the training set for the SDSS.

\begin{figure*}
  \begin{center}
    \resizebox{150mm}{!}{\includegraphics[angle=0]{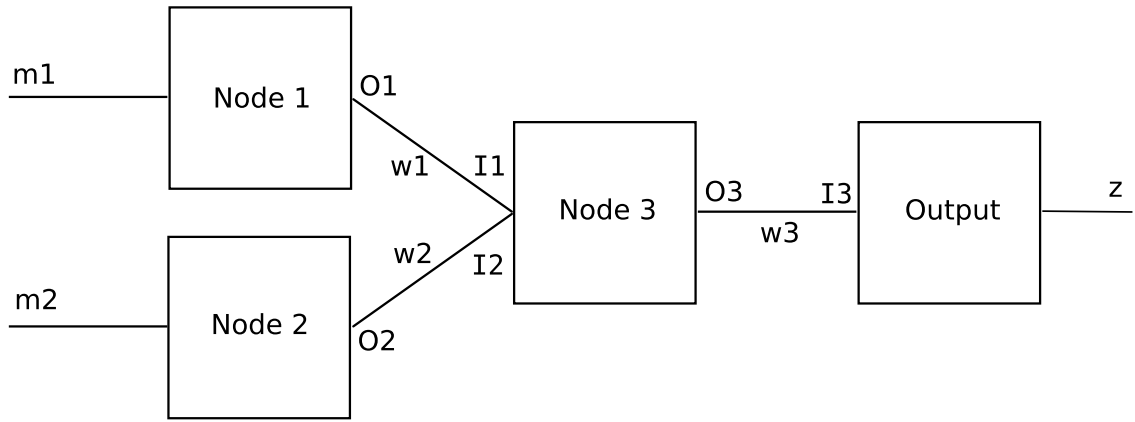}}
    \caption{A simple FFMP network with 3 layers and configuration $2:1:1$.  
The inputs are the 
two magnitudes, $m_1$ and $m_2$.
Ix denotes the input from node x, and Ox is the corresponding output of this node.
The weights $w$ associated with each connection are found by training the network
using training and validation sets (see text).}
    \label{NNsimple}
  \end{center}
\end{figure*}

We have constructed a spectroscopic sample consisting of $639,911$ 
galaxies that have SDSS photometry measurements
(counting repeats; see below) and that have 
spectroscopic redshifts measured by the SDSS or by 
other surveys, as described below. 
We imposed a magnitude limit of $r<23.0$ on the spectroscopic 
sample and applied 
additional cuts on the quality of the spectroscopic
redshifts reported by the different surveys.
Since we impose a limit of $r<22.0$ for the SDSS photometric sample,
the fainter limit chosen 
for the spectroscopic training sample accommodates the full photometric
range of interest without creating boundary effects for photo-z's of
galaxies with magnitudes near the photometric sample limit of $r = 22$. 
Each survey providing spectroscopic redshifts defines a redshift 
quality indicator; we refer the reader to the respective publications listed 
below for their precise definitions.
For each survey, we chose a redshift quality cut roughly corresponding
to 90\% redshift confidence or greater. 
The SDSS spectroscopic sample 
provides $531,672$ redshifts, principally from the MAIN and 
Luminous Red Galaxy (LRG) samples, with confidence level 
$z_{\rm conf} > 0.9$. The remaining redshifts are:
$21,123$ from the Canadian Network for Observational Cosmology 
Field Galaxy Survey \citep[CNOC2;][]{yee00},  
$1,830$ from the Canada-France Redshift Survey \citep[CFRS;][]{lil95}
with Class $> 1$,  
$31,716$ from the Deep Extragalactic Evolutionary Probe \citep[DEEP;][]{deep2}
with $q_z$ =  A or B and from DEEP2  
\citep{wei05}\footnote{{\tt http://deep.berkeley.edu/DR2/ }}
with $z_{\rm quality} \geq 3$, 
$728$ from the Team Keck Redshift Survey \citep[TKRS;][]{wir04} 
with $z_{\rm quality} > -1$, and 
$52,842$ LRGs from the 
2dF-SDSS LRG and QSO Survey 
\citep[2SLAQ;][]{can06}\footnote{{\tt http://lrg.physics.uq.edu.au/New\_dataset2/ }}
with $z_{\rm op} \geq 3$.

We positionally matched the galaxies with spectroscopic redshifts against photometric 
data in the SDSS {\tt BestRuns} CAS database, which allowed us
to match with photometric measurements in different SDSS imaging runs.
The above numbers for galaxies with redshifts count independent photometric 
measurements of the same objects due to multiple SDSS imaging of the same 
region; in particular SDSS Stripe 82 has been imaged a number of times.  
The numbers of {\em unique} galaxies used from these surveys are 
$1,435$ from CNOC2, 
$272$ from CFRS, 
$6,049$ from DEEP and DEEP2,  
$389$ from TKRS, and  
$11,426$ from 2SLAQ.
The SDSS spectroscopic samples were drawn from the SDSS primary galaxy sample and therefore are all unique.
The spectroscopic sample obtained by combining all these catalogs, 
including the repeats, was divided into two catalogs of the 
same size ($\sim 320,000$ objects each). 
One of these catalogs was taken to be 
the {\it training set} used by the photo-z and error estimators, and the other
was used as a {\it validation set} to carry out tests of photo-z
quality (see \S \ref{subsec:meth_photoz}). Our tests indicate that this 
procedure of treating different  
images of the same training/validation set galaxies as independent objects leads 
to good results, provided all the photometric measurements for a given object 
are confined to either the training set or the validation set and not mixed. By 
contrast, 
excluding such multiple images from the spectroscopic sample would result
in much smaller training and validation sets; these would be very sparse at
faint magnitudes, leading to much diminished photo-z quality there. On the other 
hand, splitting 
the repeat images of a given object between the training and validation sets 
may result in ``over-fitting'' of the derived photo-z's 
(see \S \ref{subsec:meth_photoz}). 

The $r$-magnitude and color ($g-r$ and $r-i$) 
distributions for the spectroscopic samples and for 
the photometric sample are shown in Figs. \ref{dist.sdss} and 
\ref{dist.color.sdss}. While the magnitude and color distributions of 
the combined spectroscopic sample are not 
identical to those of the photometric sample, the 
spectroscopic sample does span the  
range of apparent magnitude and color of the photometric sample.
To test the impact of having a training set that is not fully representative 
of the photometric sample, we 
divided the spectroscopic sample into smaller, alternate training and 
validation sets. For instance,
to test the effect of the training-set magnitude distribution on the
photo-z estimates, we created a training set with a flat $r$
magnitude distribution and another with an $r$ magnitude distribution similar to that 
of the  
photometric sample. Our tests indicated that the photo-z quality
is not strongly affected by the magnitude
distribution of the training set. 
The changes in the photo-z performance metrics
(the rms scatter and the 68\% CL region, defined below in 
\S \ref{res}) were smaller than $10\%$ when the training-set magnitude 
distribution was varied between these different choices. 
Since using the entire spectroscopic
sample for the training and validation sets produced marginally better results 
than all other cases tested, we have adopted this as our final choice. In addition, 
we tested the effect of the size of the training set on 
our photo-z calculations. We found that the photo-z performance metrics 
defined in \S \ref{res-photoz}
are degraded by no more than 10\% when the training set is artificially 
reduced to 10\% of its original size. Even when the training set is
reduced to $\sim 1\%$ of its original size, the photo-z performance metrics are 
degraded by less than $25\%$. This gives us confidence that 
the spectroscopic training set size used here is sufficient for extracting 
nearly optimal photo-z estimates.

\section{Methods}\label{met}

\subsection{ANN and NNP Photometric redshifts} 
\label{subsec:meth_photoz}

The ANN method that we use to estimate galaxy photo-z's is 
a general classification and interpolation tool used 
successfully in an array of fields such as hand writing recognition, 
automatic aircraft 
piloting\footnote{{\tt http://www.nasa.gov/centers/dryden/news/NewsReleases/2003/03-49.html}}, 
detecting credit card 
fraud\footnote{{\tt http://www.visa.ca/en/about/visabenefits/innovation.cfm}}, 
and extracting astronomically interesting sources in a telescope image 
\citep{bertin96}.

We use a particular type of ANN called a Feed Forward Multilayer 
Perceptron (FFMP) to map the relationship between photometric observables 
and redshifts.
An FFMP network consists of several input nodes, one or more hidden layers, 
and several output nodes, all interconnected by weighted connections 
(see Fig.~\ref{NNsimple}). 
We follow the notation of \cite{col04} and denote a network with 
$N_i$ input nodes, $N_{h_j}$ nodes in hidden layer $j$, and $N_o$ 
output nodes as $N_i:N_{h_1}:N_{h_2}:...:N_{h_m}:N_o$.
For each input object, the input photometric 
data (e.g., magnitudes, colors, concentrations, etc.) 
are fed into the input 
nodes of the FFMP, which fire signals according to the values of the 
input data.
Each node in a hidden layer receives a total input which is a weighted
sum of the outputs from the nodes in the previous layer,
 i.e., node $i$ in a hidden layer receives an input $I_i$ given by  

\begin{equation}
I_i = \sum_j w_{ij} O_j,
\end{equation}

\noindent where $O_j$ is the output of the $j^{\rm th}$ node of the previous 
layer and $w_{ij}$ is the weight of the connection between node $i$ in 
the hidden layer and node $j$ in the previous layer. 
Given the input $I_i$, the output $O_i$ of node $i$ is a function $f$ of the 
input, 

\begin{equation}
O_i=f(I_i), \label{act}
\end{equation}

\noindent where $f$ is the activation function.
Repeating this process, signals propagate up to the output nodes.
The activation function is typically a sigmoid function:

\begin{equation}
f(I_i) = \frac{1}{1 + e^{-I_i}}. \label{sigm}
\end{equation}

\noindent However, there are various alternatives, such as step 
functions and hyperbolic tangents.
\cite{van04} show that the choice of activation functions makes 
no significant difference in the result.

We use $X$:20:20:20:1 networks to estimate photo-z's, where $X$ is the
number of input photometric parameters per galaxy. 
The corresponding number of degrees of freedom (the number of weights) is
 roughly 1,000, depending on the actual value of $X$.
We use hyperbolic tangent functions as the activation function of the 
hidden layers and a linear activation function for the output layer.

Despite the occasional aura of mystery surrounding neural networks, 
an FFMP is nothing more than a complex 
mathematical function; in fact, one can always write down the analytic 
expression corresponding to a neural network function. 

Once the network configuration is specified, it can be trained to 
output an estimate of redshift given the input photometric observables.
The training process involves 
finding the set of weights $w_{ij}$ that 
minimize a score function $E$, chosen here to be 

\begin{equation}
E = \frac{1}{2}\sum_i(z_{\rm spec}^{i} - z_o^{i})^2 ~,
\label{eq:score}
\end{equation}

\noindent where $z_{\rm spec}$ is the measured spectroscopic redshift, $z_o$ is the 
output redshift of the output node, and the sum is over all galaxies 
in the training set. Note that the choice of score function is not unique, 
and different choices will in general lead to different photo-z estimates.
The minimization of this score function can be done efficiently 
because its derivatives with respect to the weights 
are available analytically.
We use a Variable Metric method as described in \cite{pre92} for the minimization.

In machine learning, over-fitting refers 
to the tendency of an algorithm with many adjustable parameters 
to fit to the noise in the training set data.
In order to avoid over-fitting, we use the technique of 
early stopping.
The spectroscopic sample is divided into two 
independent subsets, the  
{\it training} and {\it validation} sets, 
and the formal minimizations are done using the training set.
After each minimization step, the network is evaluated on the 
validation set, and 
the set of weights that performs best on the validation set 
is chosen as the final set. Another issue in machine learning is that 
minimization procedures that start at different initial choices of weights 
generally end at different local minima of the score
function. 
To reduce the chance of ending in a less-than-optimal local minimum,
we minimize five networks starting at different positions in the space of weights.
Among these, we choose the network that gives the lowest photo-z scatter 
(cf. Eq. \ref{eq:score}) 
in the validation set. 
For more details of our implementation of the ANN and its performance on
mock catalogs and real data, see \cite{cun07}.

The ANN photo-z algorithm is very flexible in the sense that it is easy 
to change the input parameters, the training set, and the network configurations.
We tried a variety of combinations of possible input photometric
observables to see their effects on photo-z quality. 
We calculated photo-z's using galaxy magnitudes, colors, and the
concentration indices for some or all of the passbands.  
The concentration index $c_i$ in passband $i$ is defined as the ratio of {\tt PetroR50} 
and {\tt PetroR90}, which are the radii that encircle 50\% and 90\% of the 
Petrosian flux, respectively. Early-type (E and S0) galaxies, with centrally 
peaked surface brightness profiles, tend to have low values of the 
concentration index, while late-type spirals, with quasi-exponential light 
profiles, typically have higher values of $c$.
Previous studies \citep{morg58,shi01,yam05,par05} have shown 
that the concentration parameter correlates well
with galaxy morphological type, and we used it to help break the 
degeneracy between redshift and galaxy type.
We present the photo-z results for different combinations of input 
parameters in \S\ref{res}.

For comparison, we also computed photo-z's for the 
validation set using another empirical method, the Nearest Neighbor 
Polynomial (NNP) technique \citep{cun07}. 
In NNP, to derive a photo-z for a galaxy in the photometric sample, 
we look for its training-set nearest neighbors in the space of 
photometric observables (magnitudes, colors, etc.).
Suppose we have $N_D$ photometric data entries for each galaxy. 
The data vector for the galaxy of interest in the photometric sample is 
denoted by $\ D^{\mu}=(D^1,D^2,...,D^{N_D})$, 
while the data vector for the $i^{\rm th}$ galaxy in the training set is 
$\ D^{\mu}_i=(D^1_i,D^2_i,...,D^{N_D}_i)$.
The distance $d_i$ between the photometric object and the $i^{\rm th}$ 
training set galaxy is defined using a flat metric in data space,

\begin{equation}
d_i^2 = \sum_{\mu=1}^{N_D} (D^{\mu} - D_{i}^{\mu})^2~. \label{nndef}
\end{equation}

\noindent The nearest neighbors are the training-set objects 
for which $d_i$ is minimum. Once the nearest neighbors for a given 
galaxy are identified,
they are used to fit the coefficients of a local, low-order polynomial relation 
between photometric observables and redshift. 
The galaxy photo-z is then obtained by applying
the derived relation to the photometric object. 

For the NNP method employed in this work, we take the 
photometric data $D^{\mu}$ in Eq.~(\ref{nndef}) 
to be the four ``adjacent'' galaxy colors $u-g, \ g-r, \ r-i, \ i-z$; we found that  
this choice produces results marginally better than using the galaxy 
magnitudes. 
We use the nearest $1000$ neighbors to fit a quadratic polynomial 
relation between redshift and the photometric data, here chosen 
to be the five magnitudes in each passband ($ugriz$) and their
corresponding concentration indices. 
We note that \cite{wan07} used a similar technique to estimate
photo-z's for a small sample of SDSS {\it spectroscopic} galaxies. 
They applied the Kernel Regression method of order 0, weighting
the training-set neighbors and computing photo-z's by using the
weighted average of the neighbors' redshifts. 
Our NNP method is closer to a Kernel Regression of order 2, since
we perform quadratic fits; however, we do not apply variable weights to the neighbors
but treat them equally in the fit.

Whereas the ANN method provides 
a single, nonlinear, global fit using the whole 
training set and applies the derived photo-z relation to all photometric objects, 
the NNP method yields a separate, linear (in parameters), local fit for 
each photometric object using its neighbors. If 
the galaxy magnitude-concentration-redshift hypersurface is a differentiable manifold, 
i.e., if it can be locally approximated by a hyperplane even though it 
is globally curved, then these two photo-z methods should be roughly 
equivalent. Indeed, as we show in \S \ref{res}, their performance is very similar.

\subsection{Photometric redshift errors}\label{meter}

We estimated photo-z errors for objects in the photometric catalog using 
the Nearest Neighbor Error (NNE) estimator \citep{oya07}. 
The NNE method is training-set based, with
a neighbor selection similar to the NNP photo-z estimator; it 
associates photo-z errors to photometric objects by considering the 
errors for objects with similar multi-band magnitudes in the 
validation set. 
We use the validation set, because the photo-z's of the training set could be
over-fit, which would result in NNE underestimating the photo-z errors.

The procedure to calculate the redshift error for a galaxy in the photometric
sample is as follows. 
We find the validation-set nearest neighbors to the galaxy of 
interest. In contrast to NNP, 
where the distance in Eq.~(\ref{nndef}) was defined in color space,
the NNE distance is defined in magnitude space, since photo-z errors 
correlate strongly with magnitude.
Since the selected nearest neighbors are in the spectroscopic sample, 
we know their photo-z errors, $\delta z = z_{\rm phot}-z_{\rm spec}$, where 
$z_{\rm phot}$ is computed using the ANN or the NNP method. 
We calculated the $68\%$ width of the $\delta z$ distribution 
for the neighbors and assigned that number as the photo-z error 
estimate for the photometric galaxy. Here we selected 
the nearest $200$ neighbors of each object to estimate its photo-z error. 
In studies of photo-z error estimators applied
to mock and real galaxy catalogs, we found that NNE  
accurately predicts the photo-z error when the training set is 
representative of the photometric sample \citep{oya07}.

\subsection{Estimating the Redshift Distribution}\label{estdist}

As we shall see in \S \ref{res-photoz}, estimates for 
galaxy photo-z's suffer from statistical biases that in general 
cannot be completely removed on an object-by-object basis. However, we 
can seek an unbiased estimate of the true redshift {\it distribution}
for the photometric sample that is independent of individual 
galaxy photo-z estimates. For some statistical applications, 
the redshift distribution of the photometric sample, as opposed 
to individual galaxy photo-z's, is all that is required.
One way to estimate this distribution is to  
assign a weight to every galaxy in the spectroscopic sample 
such that the {\it weighted} spectroscopic sample has the same 
distributions of magnitudes and colors as the photometric sample.
The $z_{\rm spec}$ distribution of this weighted spectroscopic 
sample provides an estimate of the true, underlying 
redshift distribution of the photometric sample.

The weight $W^{\alpha}$ of the $\alpha^{\rm th}$ spectroscopic 
galaxy is calculated by comparing
the local density around the galaxy in the spectroscopic sample with
the density of the corresponding region in the photometric sample.
The local density is evaluated by counting the number of 
nearest neighbors using the distance measured in the space of photometric 
observables, as in Eq.~(\ref{nndef}). We fix the number of spectroscopic 
neighbors, $N_{\rm S}$, which determines the distance $d_{\rm max}$
to the $N_{\rm S}^{\rm th}$-nearest spectroscopic neighbor. 
We then find the number of neighbors $N_{\rm P}$ in the photometric 
sample within the same distance $d_{\rm max}$ of the spectroscopic 
galaxy. Up to an arbitrary normalization factor, the weight is defined as

\begin{eqnarray}
W^{\alpha} \sim \frac{N_{\rm P} }{ N_{\rm S} } ~. 
\label{eqn:weight}
\end{eqnarray}

\noindent For our estimates, we chose $N_{\rm S}=20$, which provides a good
match of the weighted spectroscopic distributions of magnitudes 
and colors to those of the photometric sample. We note that if 
additional cuts in magnitude or color are applied to the photometric 
sample, then this procedure must be repeated for the newly selected photometric  
sample. 
More details and tests of this method and comparisons with 
other methods for estimating the
underlying redshift distribution (e.g., deconvolving the error distribution 
from the \zphot \ distribution) will be presented 
separately \citep{lim07}.


\begin{figure*}
  \begin{center}
    \begin{minipage}[t]{46mm}
      \begin{center}
      \resizebox{46mm}{!}{\includegraphics[angle=0]{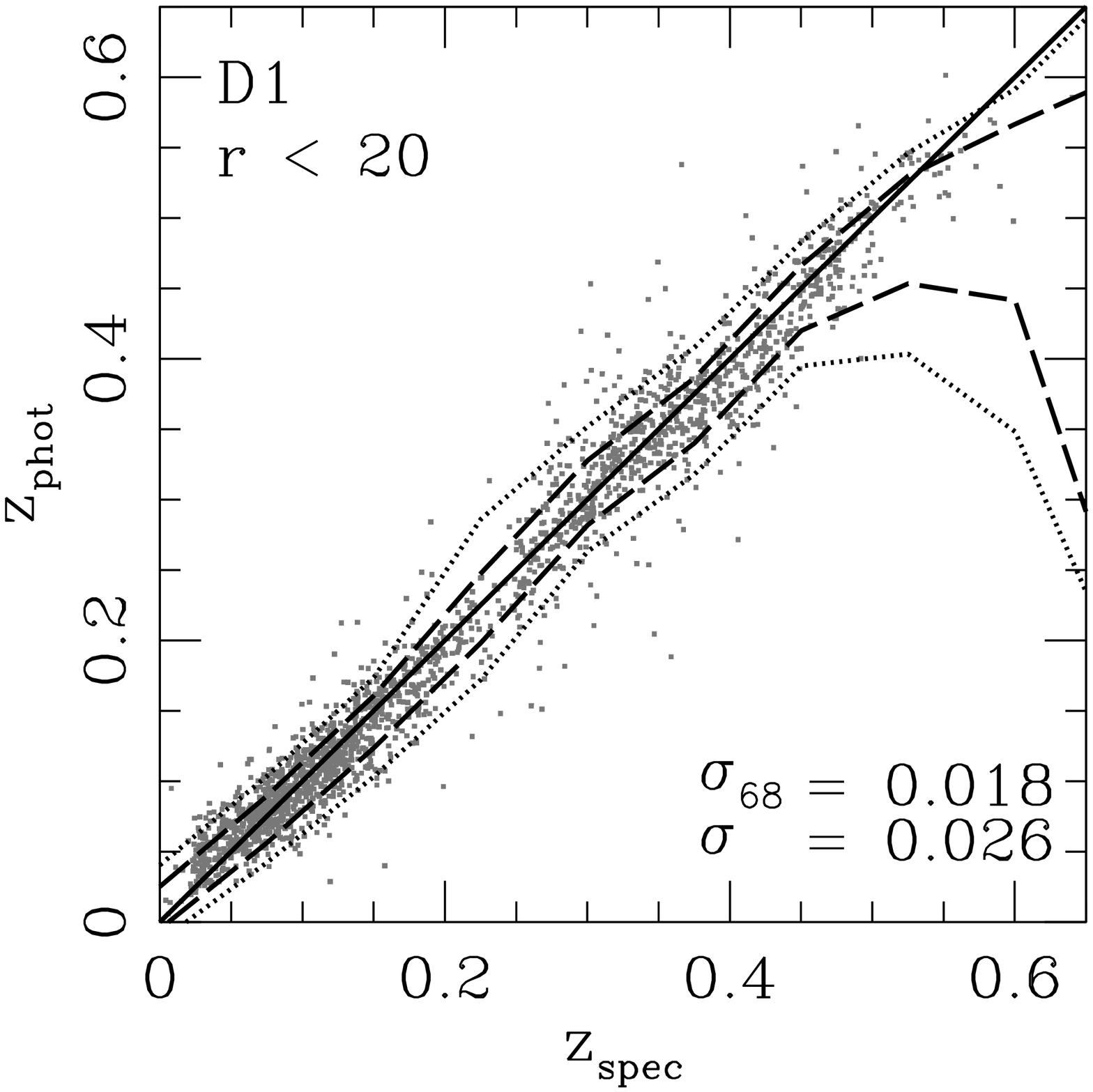}}
      \end{center}
    \end{minipage}
    \begin{minipage}[t]{46mm}
      \begin{center}
      \resizebox{46mm}{!}{\includegraphics[angle=0]{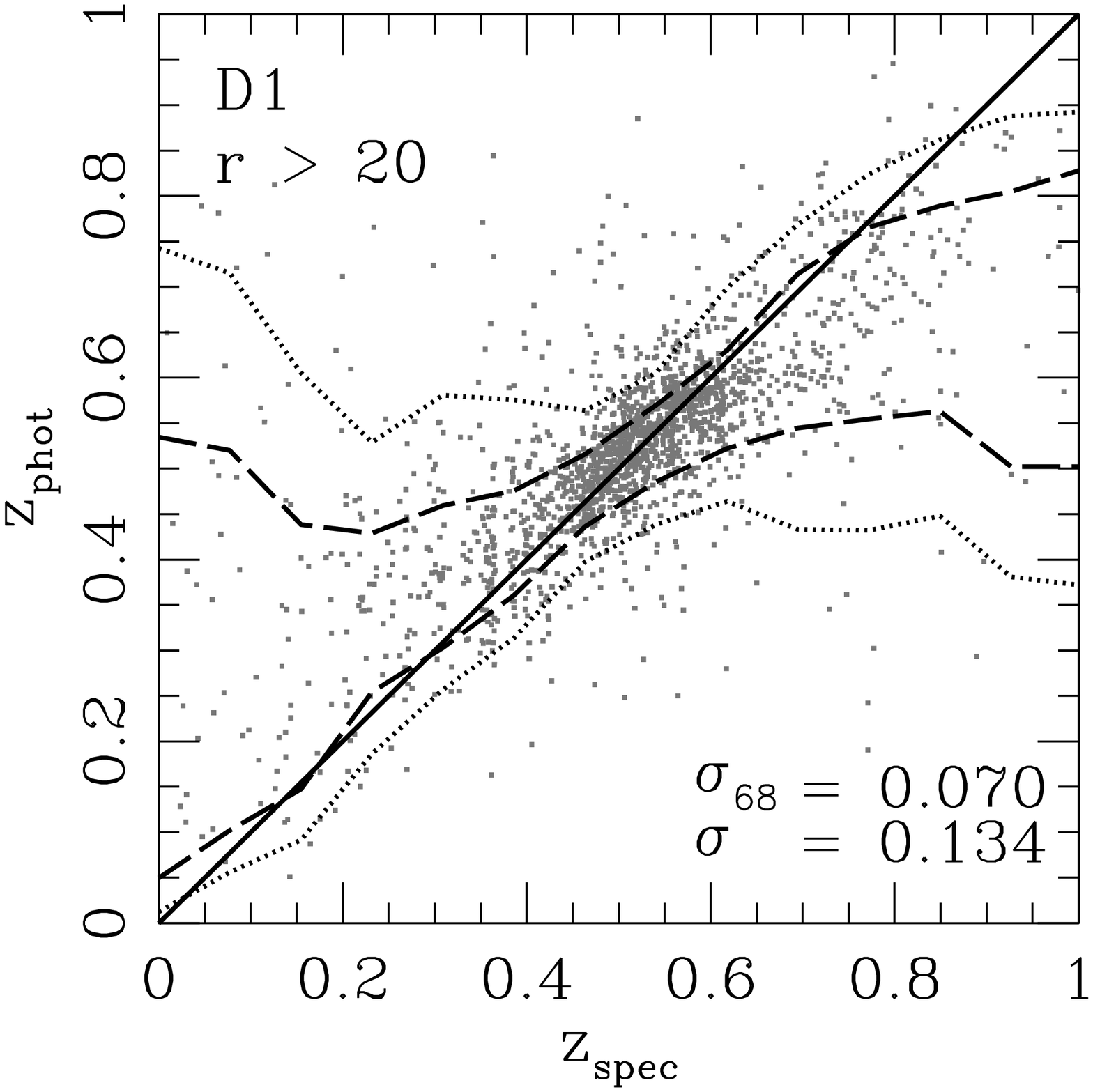}}
      \end{center}
    \end{minipage}
    \begin{minipage}[t]{46mm}
      \begin{center}
      \resizebox{46mm}{!}{\includegraphics[angle=0]{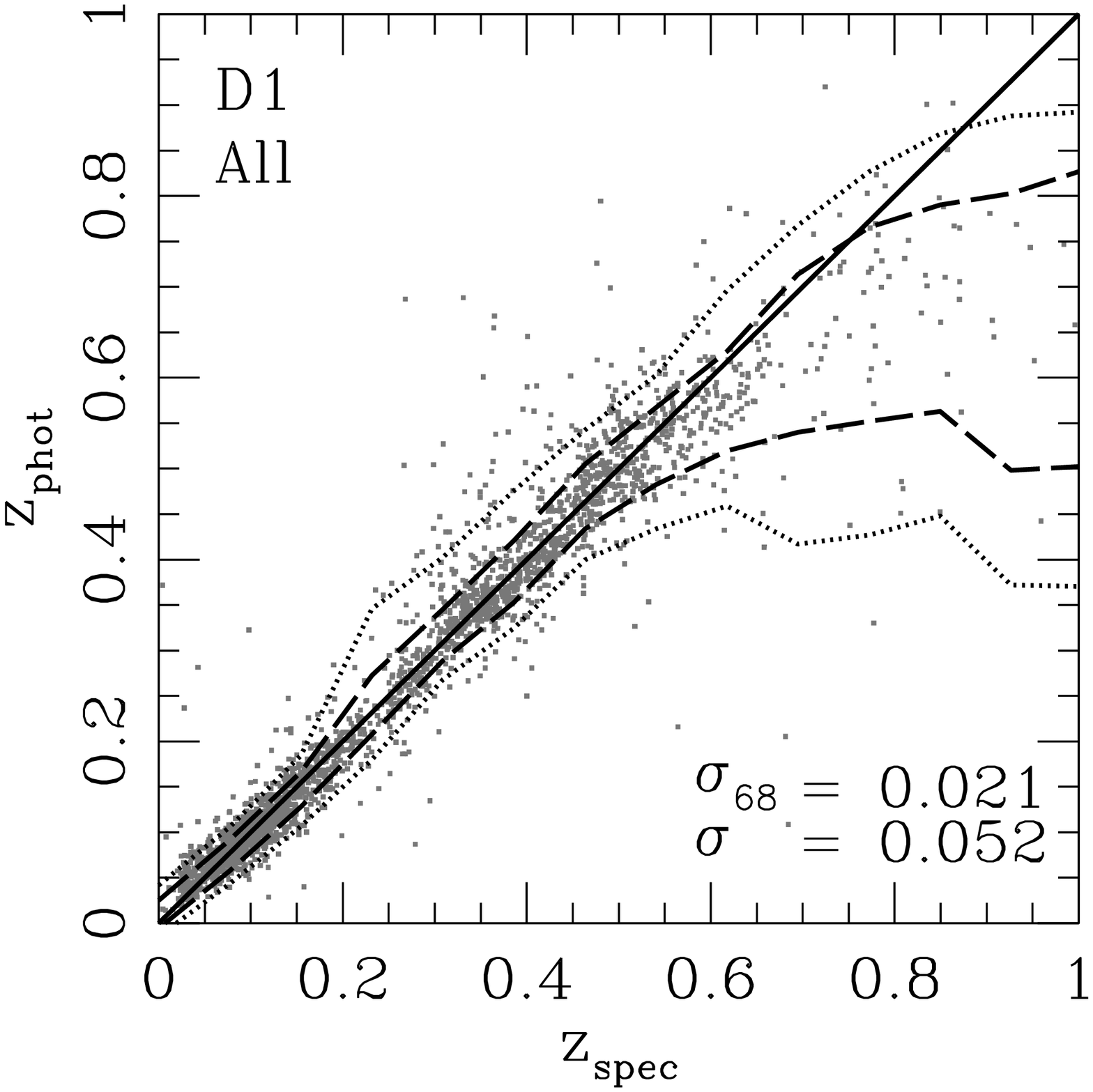}}
      \end{center}
    \end{minipage}
    \begin{minipage}[t]{46mm}
      \begin{center}
      \resizebox{46mm}{!}{\includegraphics[angle=0]{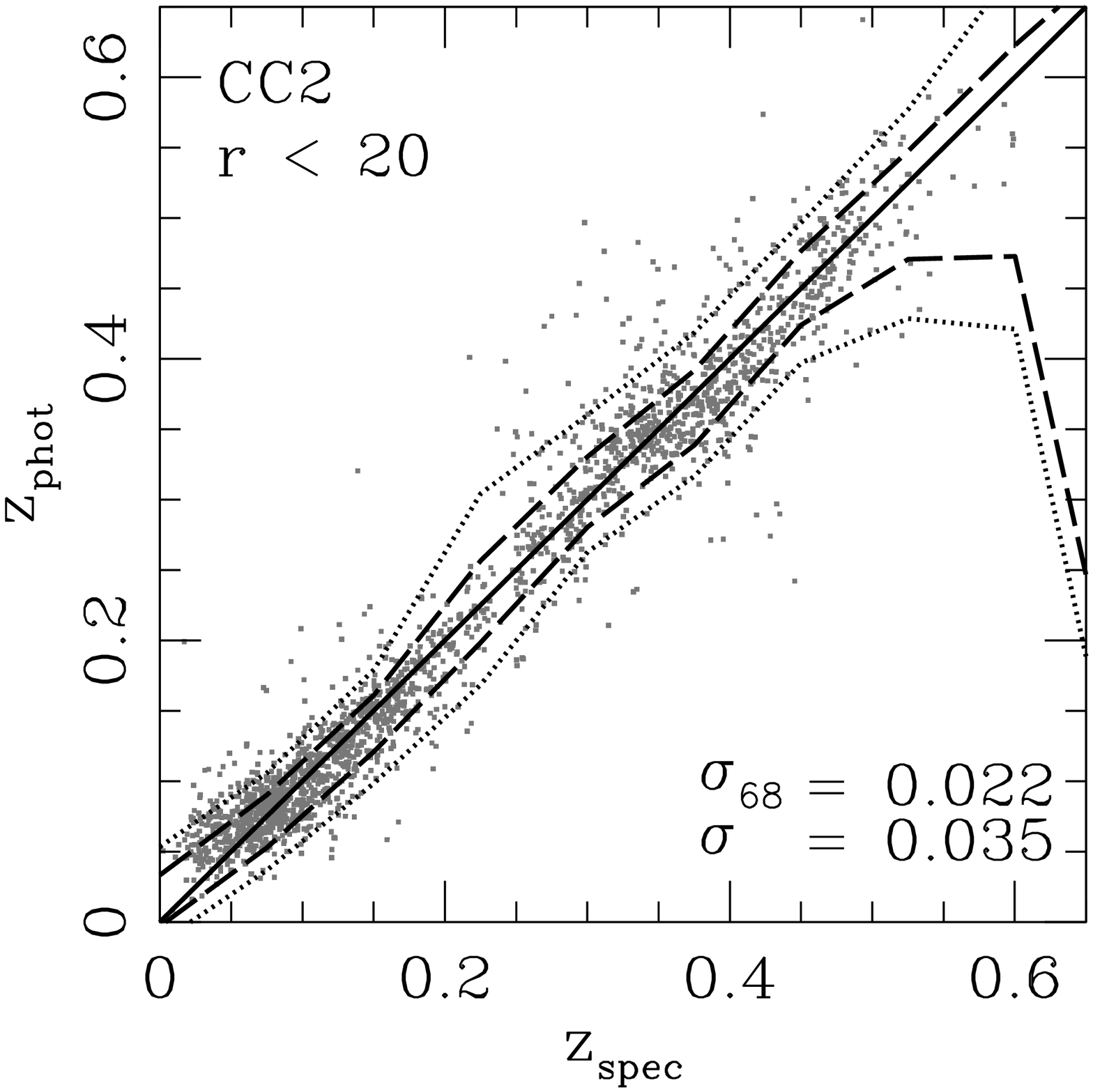}}
      \end{center}
    \end{minipage}
    \begin{minipage}[t]{46mm}
      \begin{center}
      \resizebox{46mm}{!}{\includegraphics[angle=0]{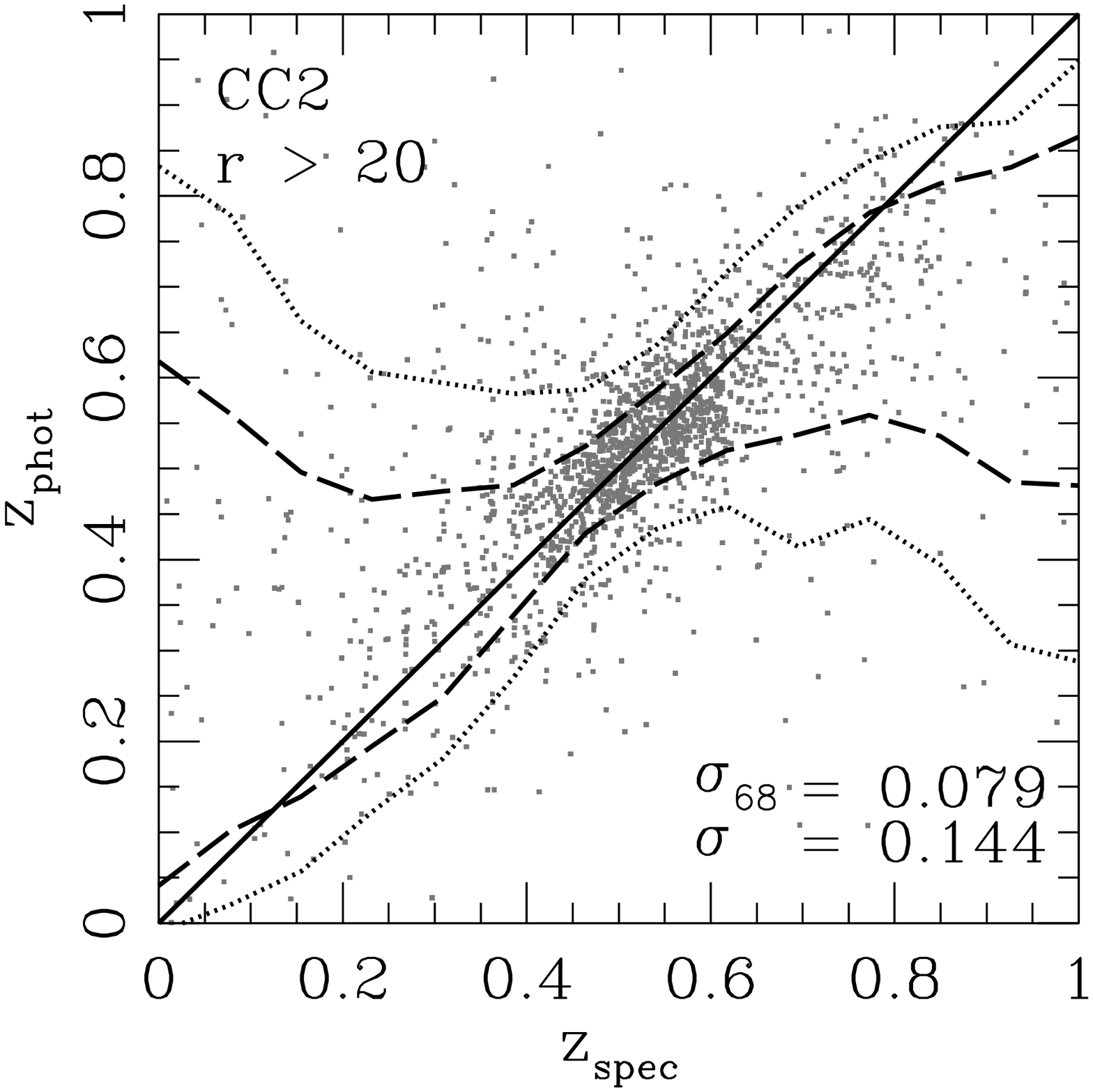}}
      \end{center}
    \end{minipage}
    \begin{minipage}[t]{46mm}
      \begin{center}
      \resizebox{46mm}{!}{\includegraphics[angle=0]{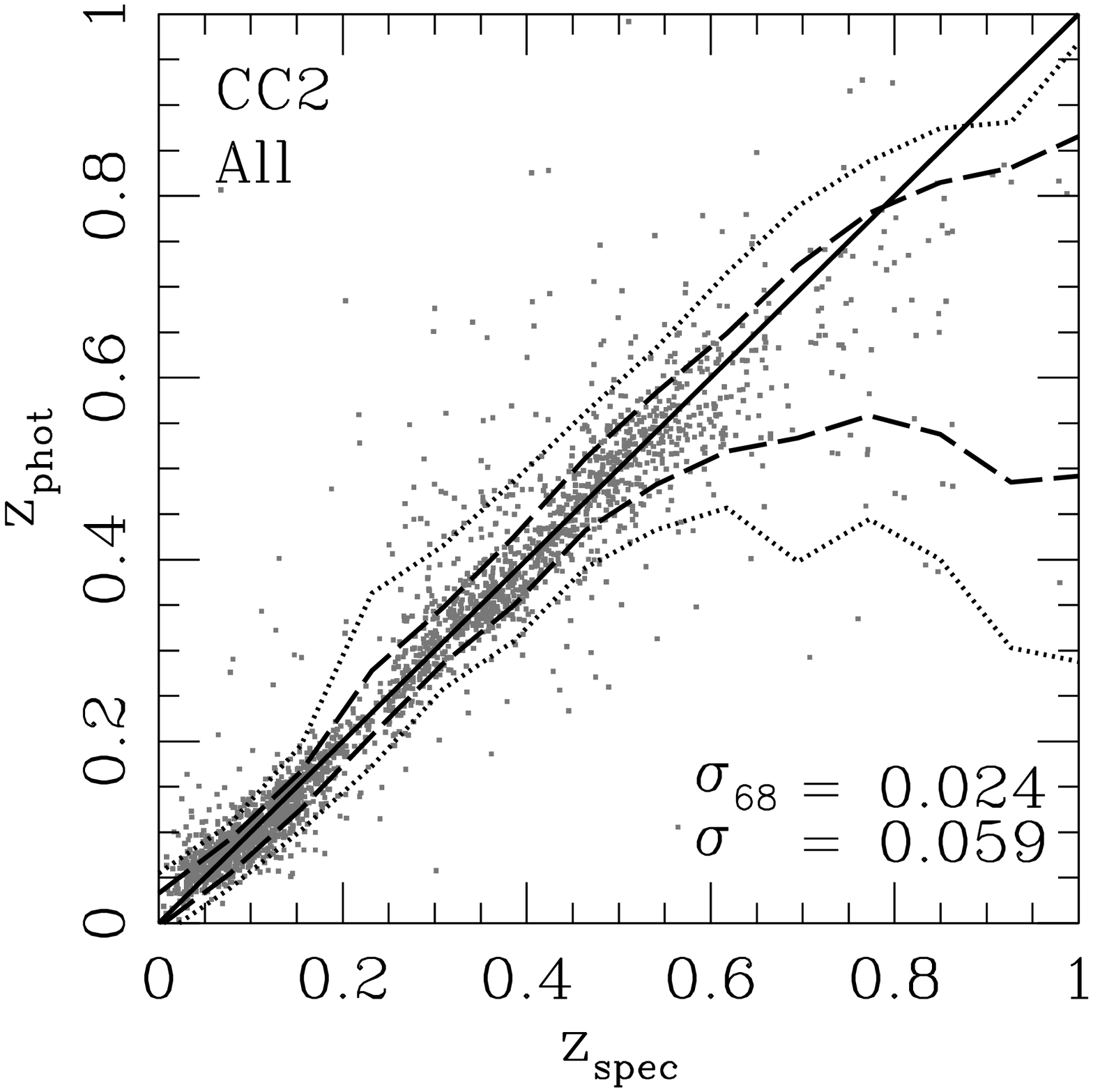}}
      \end{center}
    \end{minipage}
    \begin{minipage}[t]{46mm}
      \begin{center}
      \resizebox{46mm}{!}{\includegraphics[angle=0]{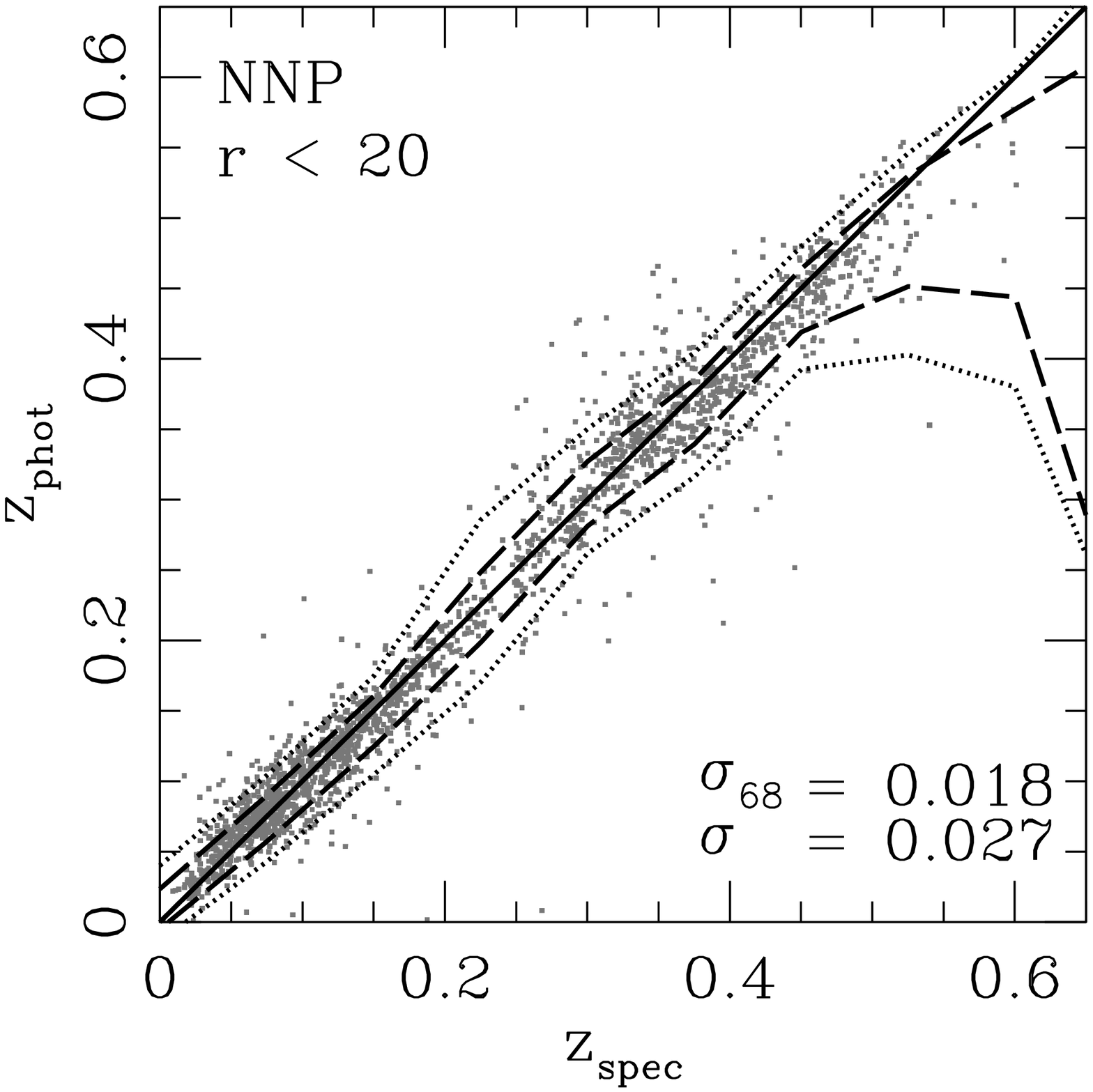}}
      \end{center}
    \end{minipage}
    \begin{minipage}[t]{46mm}
      \begin{center}
      \resizebox{46mm}{!}{\includegraphics[angle=0]{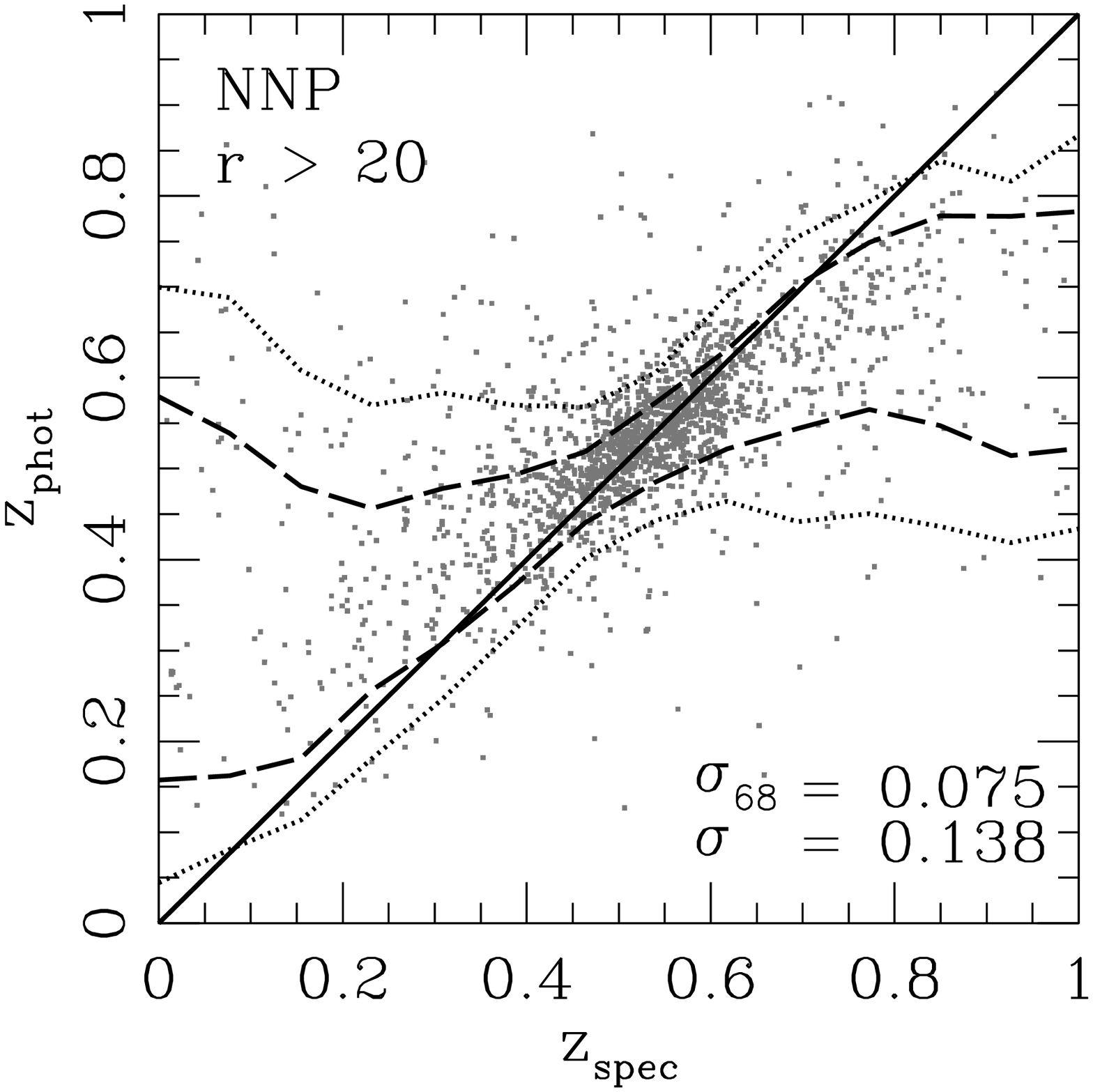}}
      \end{center}
    \end{minipage}
    \begin{minipage}[t]{46mm}
      \begin{center}
      \resizebox{46mm}{!}{\includegraphics[angle=0]{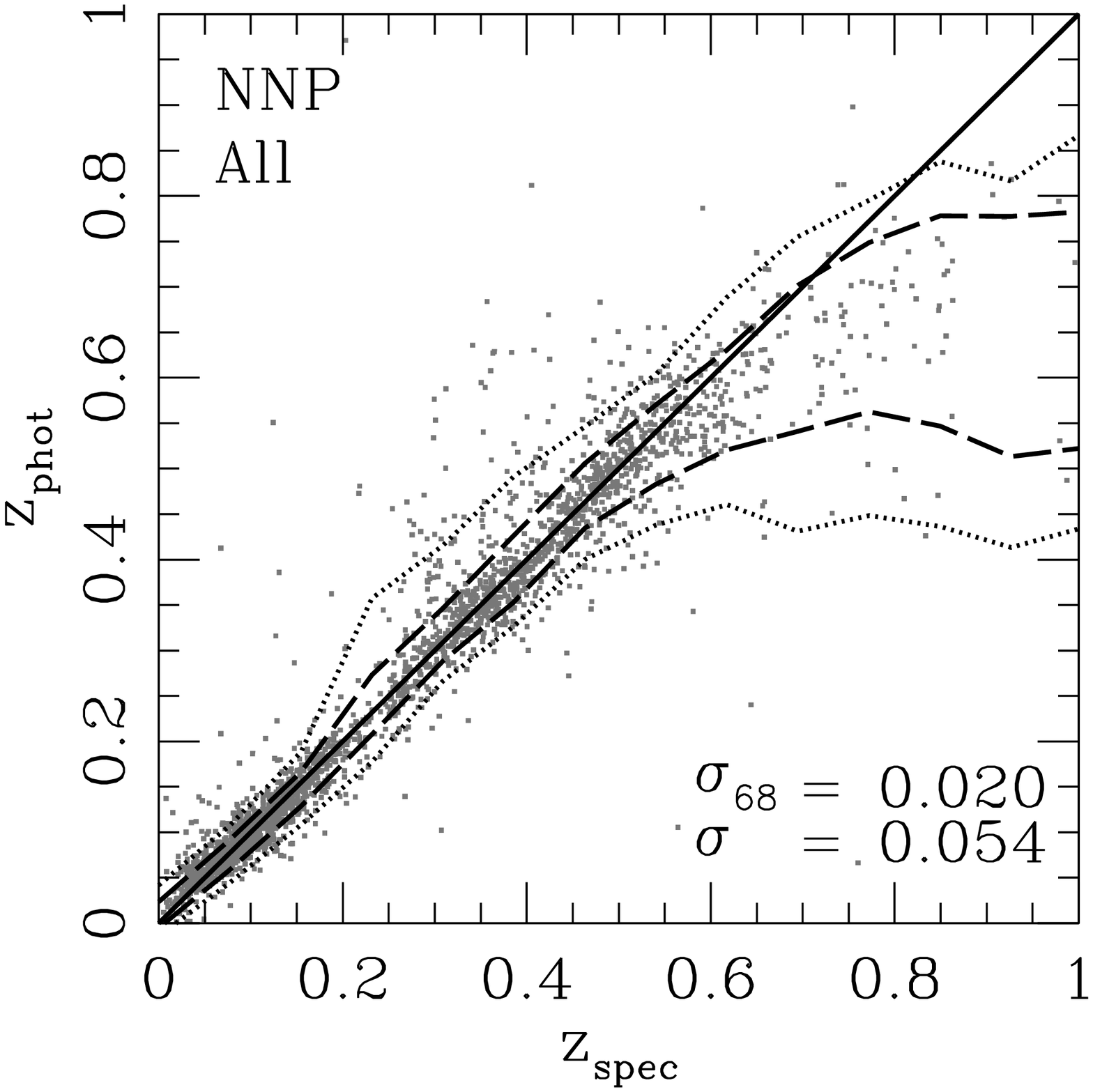}}
      \end{center}
    \end{minipage}
  \end{center}
 \caption{ $z_{\rm phot}$ versus $z_{\rm spec}$ for the validation set for 
   different ranges of $r$ magnitude and for different photo-z techniques. 
   {\it Left column:} objects with $r<20$; {\it middle column:} objects with $r>20$; 
   {\it right column:} all objects. 
   {\it Top row:} ANN case D1, where the input photometric data comprise 
   the 5 magnitudes ($ugriz$) and the 5 concentration parameters, and the training 
   is split into 5 bins of $r$ magnitude
   {\it Middle row:} ANN case CC2, where the input data are 
   the 4 colors $u-g$, $g-r$, $r-i$, $i-z$, and 3 concentration parameters $c_gc_rc_i$. 
   {\it Bottom row:} results for the NNP method, where the input data are 
   the 5 magnitudes and 5 concentration parameters. 
   In all cases, the photo-z methods 
   used a training set with $\sim 320,000$ objects, and the derived solutions were
   applied to an independent validation set with $\sim 309,000$ objects and
   $r < 22$, reflecting the magnitude limit of the photometric sample. 
   The solid line in each panel indicates $z_{\rm phot}=z_{\rm spec}$; the 
   dashed and dotted lines show the 68\% and 95\% confidence regions as a function
   of $z_{\rm spec}$.  
   The points display results for a random $10\%$ subset of the validation set in 
   each magnitude range. 
 }
\label{zpzs_valid_all}
\end{figure*}

\section{Results} \label{res}

\subsection{Photometric redshifts}
\label{res-photoz}

The photo-z precision (variance) and accuracy (bias) are 
limited by a number of factors. There are 
intrinsic degeneracies in 
magnitude-redshift space: low-luminosity, intrinsically red galaxies at low 
redshift can have apparent magnitudes similar to those of high-luminosity, 
intrinsically blue galaxies at high redshift. 
This natural degeneracy is amplified by  
photometric errors, since magnitude uncertainties 
propagate to photo-z errors. 
In addition to these observational limitations, which are 
determined by the photometric precision and the number of passbands of a survey, 
the photo-z estimator itself may have inherent limitations. For example, 
for training set methods, the size and representativeness of the training 
set are important factors, as are the number of parameters or weights in 
the fitting functions.

To test the quality of the photo-z estimates, 
we use four photo-z performance metrics. 
The first two metrics are the photo-z bias, $z_{\rm bias}$, and the photo-z {\it rms} 
scatter, $\sigma$, both averaged over all $N$ objects in the validation 
set, defined by 
  
\begin{eqnarray}
z_{\rm bias}&=&\frac{1}{N}\sum_{i=1}^{N}\left( z_{\rm phot}^{i}-z_{\rm spec}^{i}\right) ~, \\
\sigma^2&=&\frac{1}{N}\sum_{i=1}^{N}\left(z_{\rm phot}^{i}-z_{\rm spec}^{i}\right)^2 ~.
\end{eqnarray}

\noindent The third performance metric, denoted by $\sigma_{68}$, is 
the range containing $68\%$ of the validation set objects in the distribution of 
$\delta z = z_{\rm phot}-z_{\rm spec}$. This metric is useful because 
the probability distribution function 
$P(\delta z)$ is in general non-Gaussian and asymmetric (for a Gaussian 
distribution, $\sigma$ and $\sigma_{68}$ coincide). Explicitly, $\sigma_{68}$ is 
defined by the value of $|z_{\rm phot} - z_{\rm spec}|$ such that 68\% of the objects have $|z_{\rm phot} - z_{\rm spec}| < \sigma_{68}$.
We also use the $95\%$ region $\sigma_{95}$, defined similarly. 
In addition to these global metrics, we also define local versions of them 
in bins of redshift or magnitude.

\begin{deluxetable}{llcc}
\tablewidth{0pt}
\tablecaption{Summary of ANN cases}
\startdata
\hline
\hline
\multicolumn{1}{c}{Case} & \multicolumn{1}{c}{Inputs/Description} & \multicolumn{1}{c}{$\sigma$} & \multicolumn{1}{c}{$\sigma_{68}$}\\
\hline
O1& $ugriz$                                    &0.0525 & 0.0229\\ 
C1& $ugriz$ + $c_uc_gc_rc_ic_z$                &0.0519 & 0.0224\\
D1& $ugriz$ + $c_uc_gc_rc_ic_z$. Split training&0.0519 & 0.0209\\ 
CC1&$u-g$, $g-r$, $r-i$, $i-z$                 &0.0668 & 0.0272\\  
CC2&$u-g$, $g-r$, $r-i$, $i-z$ + $c_gc_rc_i$   &0.0593 & 0.0245\\ 
\enddata
\label{table:method}
\tablecomments{Photo-z performance metrics $\sigma$ and $\sigma_{68}$ 
for the validation set using different input parameters 
(magnitudes, colors, and concentration indices) and training procedures.}
\end{deluxetable}

To search for an optimal photo-z estimator, we computed 
photo-z's using the ANN method with  
different combinations of input photometric observables. Five of 
these combinations are listed in Table \ref{table:method}. 
In the first case, dubbed O1, the training and photo-z estimation
are carried out using only the five magnitudes $ugriz$. In case C1, 
we use the five magnitudes and the five concentration indices 
$c_uc_gc_rc_ic_z$ as the input parameters. In case CC1, we 
use only the four colors
$u-g$, $g-r$, $r-i$, and $i-z$. In case CC2, we combine the 
four colors with 
the concentration indices $c_gc_rc_i$ in the $gri$ filters.
Finally, in case D1, we use the $ugriz$ magnitudes
and the $c_uc_gc_rc_ic_z$ concentration indices, but we split the 
training set and the photometric sample into 5 bins of $r$ magnitude and 
perform separate ANN fits in each bin.   
In all five cases, we use an ANN with three hidden layers and tune 
the number of hidden nodes to keep the total 
number of degrees of freedom of the network roughly the same for all cases.

Table~\ref{table:method} provides a summary of the performance results of the 
different ANN cases.  
We find that using concentration indices in addition to magnitudes 
(C1 vs. O1) helps break some degeneracies and reduces the 
photo-z scatter by a few percent.  
Using only colors (CC1) degrades the photo-z performance by as much as 20\%, 
mostly because the degeneracy between intrinsically red, nearby galaxies 
and intrinsically blue, distant galaxies (with red observed colors)
cannot be broken.  
Adding concentration indices to color-only training (CC2) 
helps break such a degeneracy, because the concentration index correlates
with galaxy type and hence intrinsic color. Of the five, 
case CC2 also yields the most realistic photometric redshift
distribution for the photometric sample (see \S \ref{subsec:red_dist}).
Finally, splitting the training set and photometric sample into 
magnitude bins (D1) produces
results with the best performance metrics ($\sigma$ and $\sigma_{68}$) of
all the ANN cases we have tested. 
We choose D1 and CC2 as the best ANN cases and describe their
results in more detail below; their outputs for the photometric sample 
are included in the public DR6 database.

\begin{figure}
  \resizebox{85mm}{!}{\includegraphics[angle=0]{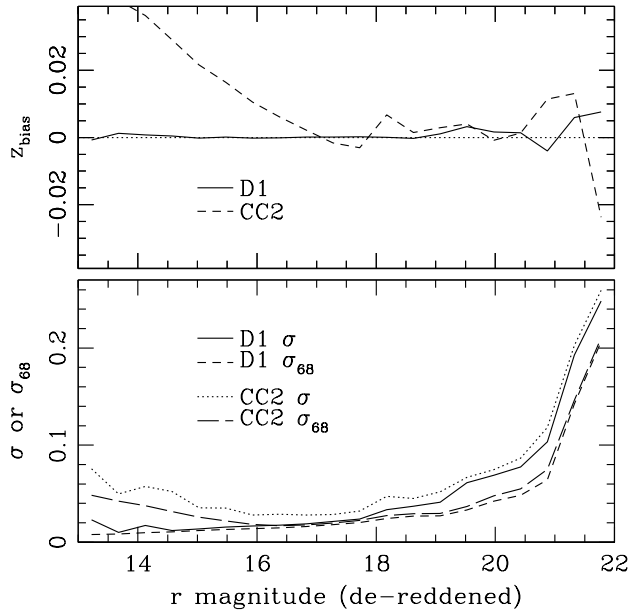}}
 \caption{The performance metrics 
   $z_{\rm bias}$, $\sigma$, and $\sigma_{68}$ for the ANN D1 and CC2 
   validation sets are shown 
   as a function of $r$ magnitude. 
   CC2 performs relatively poorly for bright objects ($r < 16$), where the color-redshift
   relation is contaminated by faint objects with similar colors.  In D1,
   this problem is alleviated by the effective magnitude prior imposed by
   the training set. At faint magnitudes, the performance degrades as the photometric 
   errors increase.
 }
 \label{plot:statvsm}
\end{figure}

In Fig.~\ref{zpzs_valid_all}, we plot photometric redshift, \zphot, 
for all objects in the validation set vs. true
spectroscopic redshift, \zspec, for the different photo-z methods  
and cases and in different ranges of $r$ magnitude. 
The top row shows results for ANN case D1, the middle row shows 
the performance of ANN case CC2, and the bottom row shows results for 
the NNP method using magnitudes and concentration indices as the input 
parameters. In each panel, 
the values of the corresponding global
photo-z performance metrics $\sigma$ and $\sigma_{68}$ are shown. 
The redshift bias $z_{\rm bias}$ is typically much smaller than $\sigma$ or 
$\sigma_{68}$, since the photo-z methods are designed to minimize it (see 
Fig. \ref{plot:statvsm}). In each panel of Fig. \ref{zpzs_valid_all}, 
the solid line traces 
$z_{\rm phot}=z_{\rm spec}$, i.e., the line 
for a perfect photo-z estimator. 
The dashed and dotted lines show the corresponding $68\%$ and $95\%$ regions, 
defined as above but in $z_{\rm spec}$ bins. Although  
each photo-z method probes the 
hypersurface defined by the photometric observables and redshift in a different 
way, 
they produce very similar results, suggesting that our results are 
limited not by the photo-z technique employed but by the 
intrinsic degeneracies in magnitude-concentration-redshift space and 
by the photometric errors.

\begin{figure*}
  \begin{center}
    \begin{minipage}[t]{81mm}
      \begin{center}
      \resizebox{81mm}{!}{\includegraphics[angle=0]{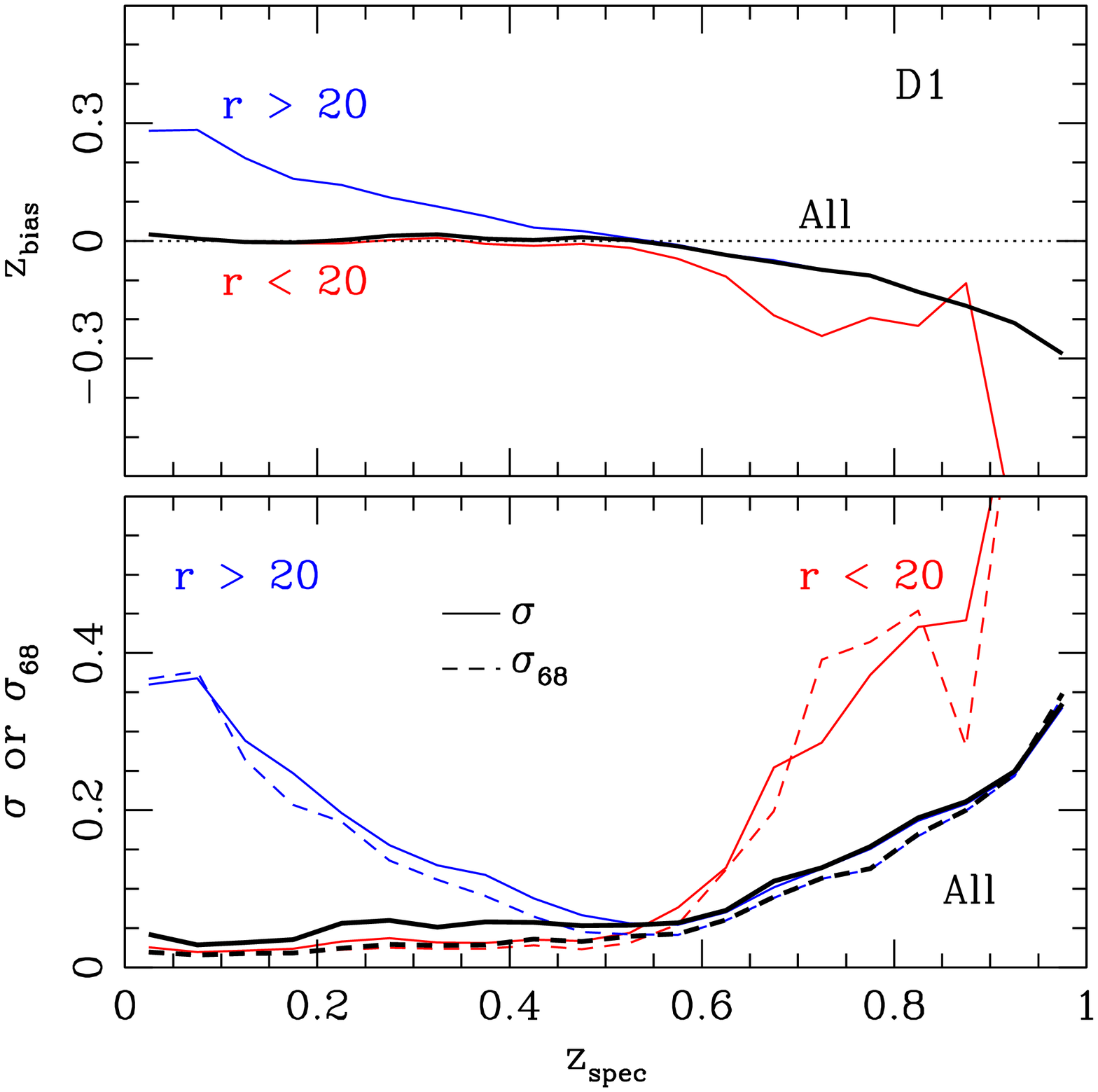}}
      \end{center}
    \end{minipage}
    \begin{minipage}[t]{81mm}
      \begin{center}
      \resizebox{81mm}{!}{\includegraphics[angle=0]{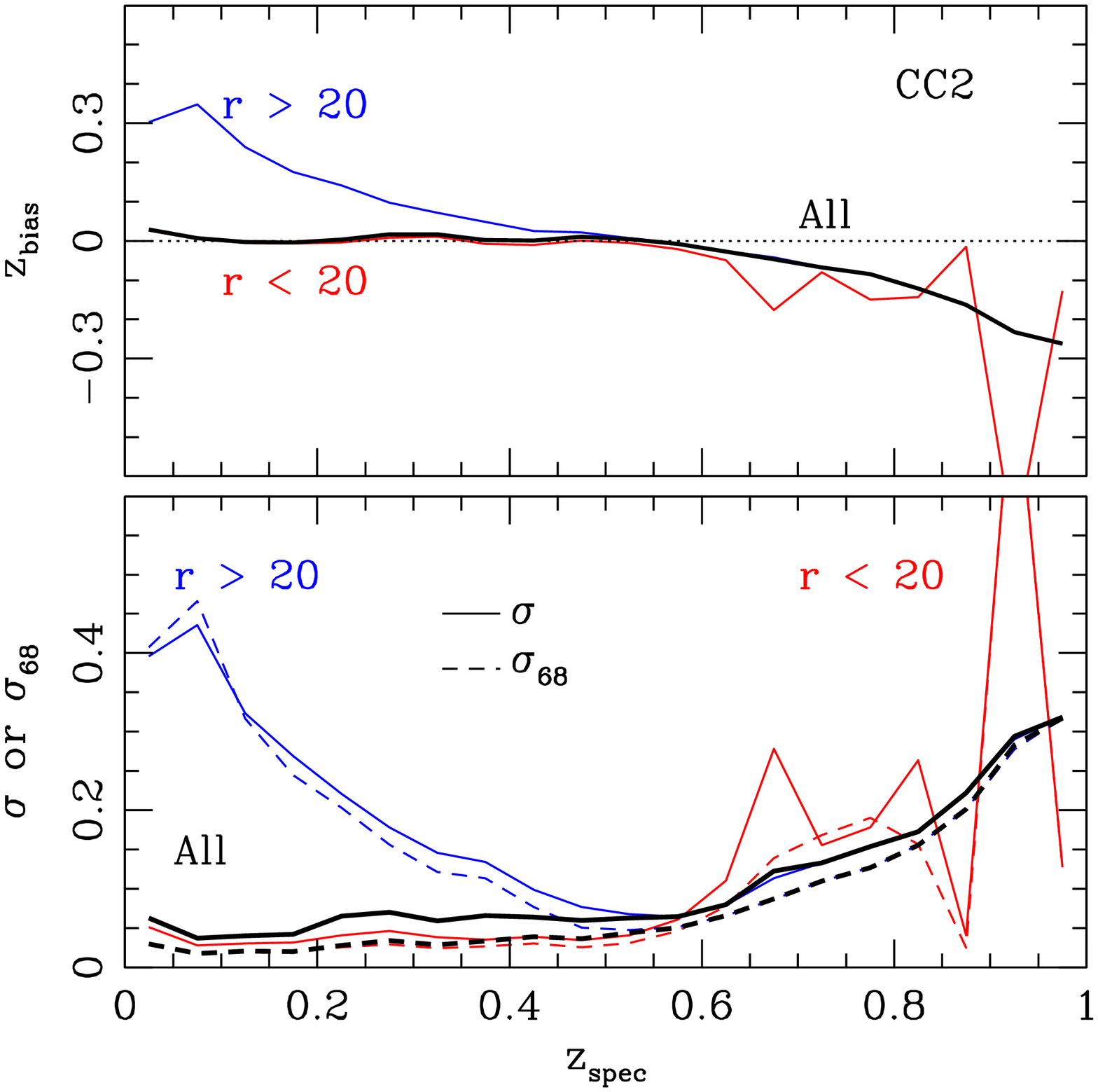}}
      \end{center}
    \end{minipage}
  \end{center}
  \caption{Performance metrics 
$z_{\rm bias}$, $\sigma$, and $\sigma_{68}$ for the ANN D1 and CC2 validation sets 
    are shown as a function of $z_{\rm spec}$ for $r<20$ and $r>20$. 
    The increased scatter for  objects with $z > 0.6$ is due to 
    the 4000 \AA \ break shifting out of the $r$ passband at
    around $z = 0.7$; beyond that redshift, the estimator effectively relies
    on only two passbands ($i$ and $z$) to determine the photo-z's. Note that 
    faint objects ($r > 20$) have worse scatter at low redshifts for
    both cases.  This is likely due to the fact that the faint, low-redshift
    objects in the validation set are predominantly blue 
    dwarf or irregular galaxies that do not have  
    strong 4000 \AA \ breaks; in this case, the photo-z estimator must rely on less
    pronounced spectral features, resulting in larger photo-z scatter.
  }
  \label{plot:statvsz}
\end{figure*}

\begin{figure*}
  \begin{center}
    \begin{minipage}[t]{81mm}
      \begin{center}
	 \resizebox{81mm}{!}{\includegraphics[angle=0]{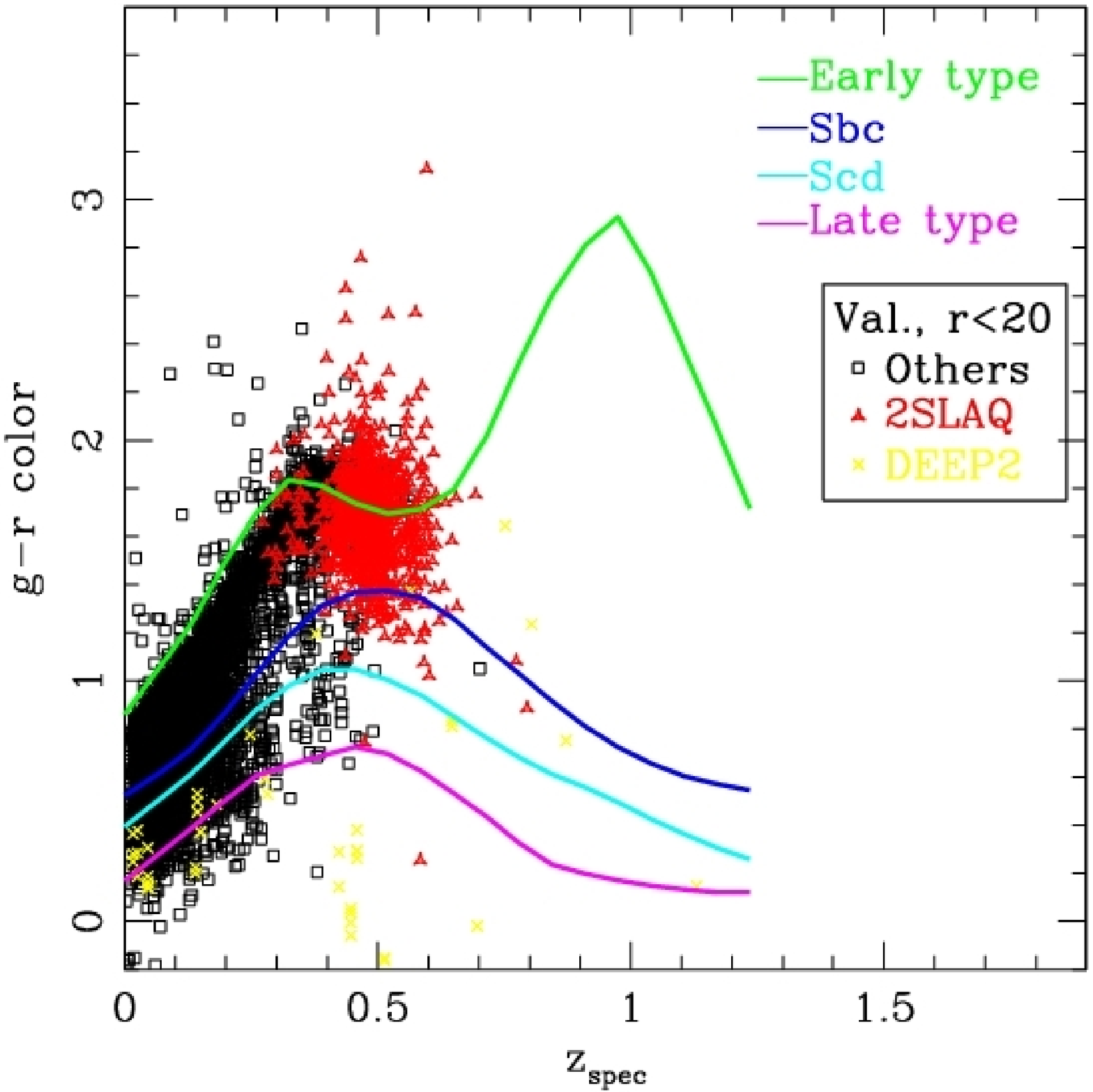}}
      \end{center}
    \end{minipage}
    \begin{minipage}[t]{81mm}
      \begin{center}
	 \resizebox{81mm}{!}{\includegraphics[angle=0]{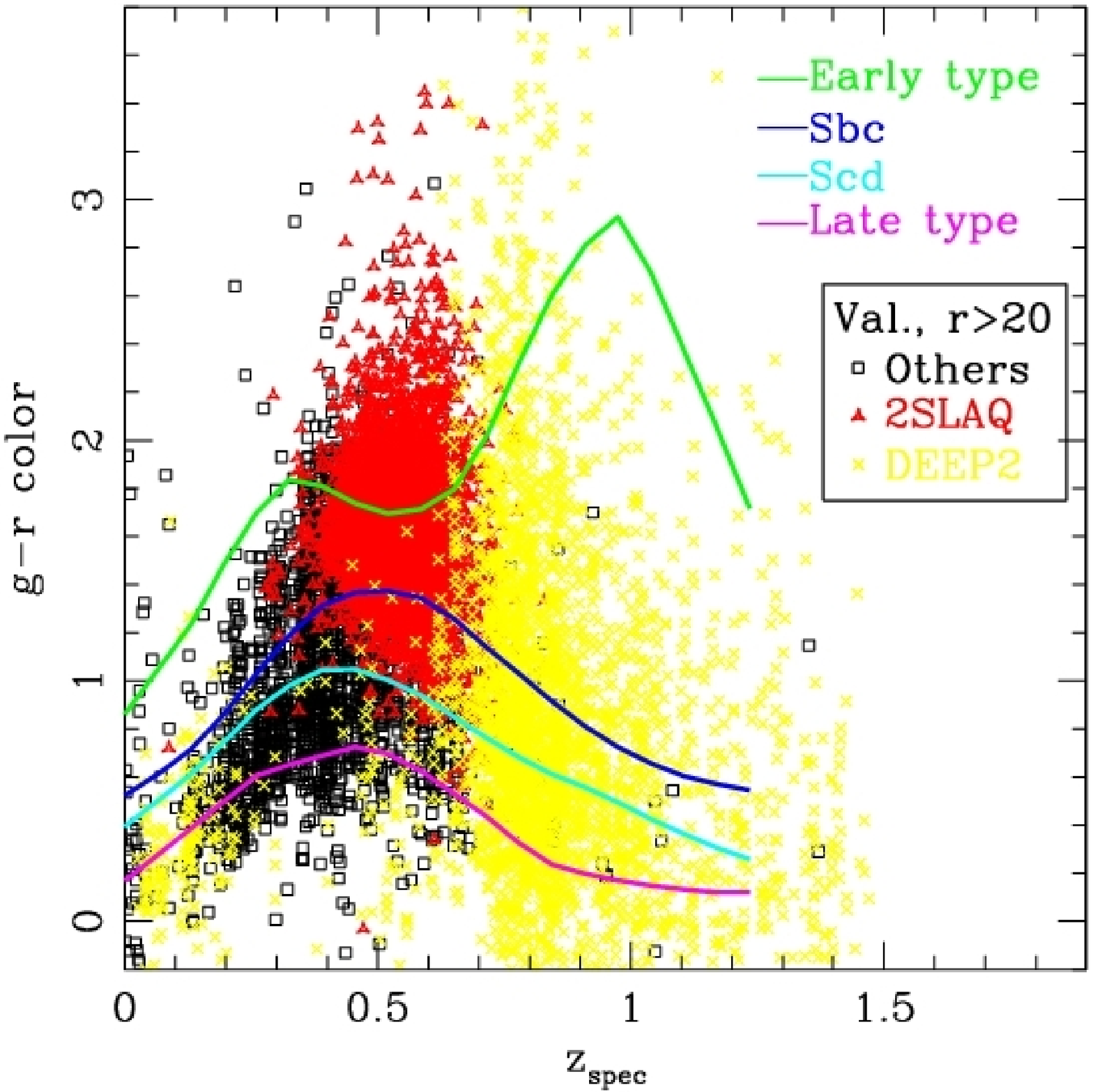}}
      \end{center}
    \end{minipage}
  \end{center}
  \caption{
    $g-r$ color vs spectroscopic redshift for galaxies in the 
    validation set: {\it left panel:} galaxies with $r<20$; {\it right panel:} 
    galaxies with $r>20$. The solid curves show expected color-redshift relations of
    galaxies with different SED types, calculated using the \cite{col80} 
    spectral templates. The different 
    colors (shades of grey) 
    indicate galaxies from the different spectroscopic surveys contributing 
    to the validation set. The 2SLAQ objects, denoted by red triangles, were 
    selected to be mostly early-type galaxies. They are
    responsible for the minimum in $\sigma$ vs. $z_{spec}$  
    for the $r>20$ subsample in Fig. \ref{plot:statvsz}.
  }
  \label{plot:grvsz}
\end{figure*}

In Figs. \ref{plot:statvsm} and \ref{plot:statvsz}, we show the performance 
metrics
$z_{\rm bias}$, $\sigma$, and $\sigma_{68}$ as a function of $r$ magnitude 
and $z_{\rm spec}$ for the validation set for the two preferred ANN cases.
We see that the photo-z precision degrades considerably
for objects with $r > 20$. 
This increased scatter is expected, since the relative photometric errors 
increase as the nominal detection limit of the SDSS photometry is approached 
(see Table \ref{propphot}). While the bias for CC2 increases at $r<17$,  
we note that the fraction of objects in the photometric sample which are 
that bright is very small.
As a function of redshift, $\sigma$ and $\sigma_{68}$ increase dramatically
beyond $z \sim 0.6$ 
for the validation set.
For the $r < 20$ part of the sample, the number of spectroscopic objects with 
$z > 0.6$ is simply too small
to characterize the redshift-magnitude surface, as shown in 
the left panel of Fig. \ref{plot:grvsz}. For the 
faint objects ($r > 20$), the scatter is low for $z$ between 0.4 and 
0.6 and increases outside of that range. 
It's important to note that the photo-z performance metrics were 
calculated independently of spectral type. 
Since the the neural network and the training set were not optimized 
for any specific galaxy population (e.g., galaxies in clusters) it is possible
that certain galaxy types may have photo-z's with worse (or better!) 
biases and dispersion.

In Figure~\ref{plot:grvsz}, we plot $g-r$ color versus spectroscopic 
redshift for the validation set for both bright ($r<20$) and faint ($r>20$) galaxies.
The 2SLAQ and DEEP2 galaxies are highlighted by different 
colors (shades of grey),
and the expected color-redshift relations for the four spectral templates from 
\cite{col80}
(from early to late types) are indicated by the solid lines.
We see that for the faint sample,  in the range $0.4 < z < 0.6$, the galaxies
come mostly from the 2SLAQ survey, which used 
specific color cuts to select early-type galaxies at
$z\sim0.5$.  Because early-type galaxies have a well-defined 
4000 \AA \ break feature, their photo-z's are well determined and 
their photo-z scatter is low.
Outside of the range $0.4 < z < 0.6$, the validation set at faint magnitudes 
is dominated by bluer galaxies
that do not have strong, broad spectral features, resulting in the 
larger photo-z scatter seen in Fig. \ref{plot:statvsz}.

Fig.~\ref{plot:statvsz} shows that the common assumption that the 
photo-z scatter
scales as $(1+z)$ is not consistent with our estimates for the SDSS sample.
The functional form of the scatter versus redshift depends 
strongly on the underlying galaxy type distribution.

\subsection{Redshift Distributions} 
\label{subsec:red_dist}

So far, we have considered the scatter and bias of photo-z estimates. 
As discussed in \S \ref{estdist}, it is also of interest to consider 
the predicted photo-z distribution as a whole. Different photo-z estimators 
may achieve similar values for the metrics $z_{\rm bias}$, $\sigma$, and $\sigma_{68}$, 
but predict different forms for the photo-z distribution of the photometric 
sample. As we shall see, this is the case with the two ANN cases D1 and CC2.
We therefore define two additional performance metrics to quantify the 
quality of the predicted photo-z distribution.
The first metric, $\sigma_{\rm dist}$, measures the {\it rms} difference between 
the binned $z_{\rm phot}$ and $z_{\rm spec}$ distributions of the validation set,

\begin{eqnarray}
\sigma^2_{\rm dist}&=&\frac{1}{N_{\rm bin}}\sum_{i=1}^{N_{\rm bin}}\left(P_{\rm phot}^{i}-P_{\rm spec}^{i}\right)^2, 
\end{eqnarray}

\noindent where $P_{\rm phot}^{i}$ is the height of the 
$i^{\rm th}$ redshift bin of the $z_{\rm phot}$ distribution,  
$P_{\rm spec}^{i}$ is the height of the same redshift
bin of the $z_{\rm spec}$ distribution, and $N_{\rm bin}$ is the total number 
of redshift bins used.  
Here we use $N_{\rm bin}=120$ equally spaced redshift bins running 
from $z=0$ to $z=1.2$.

The second redshift distribution 
metric we employ is the KS statistic $D$, the 
maximum value of the absolute difference between the two ($z_{\rm phot}$ and 
$z_{\rm spec}$) cumulative 
redshift distribution
functions. An advantage of the KS statistic is that it does not require 
binning the data in redshift. However, our 
use of the KS statistic to quantify the difference between the $z_{\rm phot}$
and $z_{\rm spec}$ distributions of the validation set likely does 
not adhere to formal statistical practice, 
since it turn outs that the probability for the KS statistic for both cases we consider 
is very close to zero \citep{pre92}.

Table \ref{table_sigdist_ks} shows the values of 
$\sigma_{\rm dist}$ and of the KS statistic $D$ for the validation set for the 
D1 and CC2 ANN photo-z's, for different ranges of $r$ magnitude. 
Although the CC2 photo-z distribution is 
a worse overall match to the $z_{\rm spec}$ distribution for the 
validation set, it works better than D1 for $r>18$. 
Since the photometric sample 
is dominated by objects at $r>20$ (see Fig. \ref{dist.sdss}), 
these results suggest that CC2 should do a better job in 
estimating the redshift distribution of the photometric sample, 
even though D1 performs better by the standards of $z_{\rm bias}$ and 
$\sigma$.

\begin{deluxetable}{cc|cc|ccc}
\tablewidth{0pt}
\tablecaption{$\sigma_{\rm dist}$ and KS statistic for Redshift distribution}
\startdata
\hline
\hline
\multicolumn{1}{c}{} & \multicolumn{1}{c|}{} & \multicolumn{2}{c|}{$\sigma_{\rm dist}$} & \multicolumn{2}{c}{KS statistic}\\
\hline
\multicolumn{1}{c}{} & \multicolumn{1}{c|}{$r$-mag bin} & \multicolumn{1}{c}{CC2} & \multicolumn{1}{c|}{D1} & \multicolumn{1}{c}{CC2} & \multicolumn{1}{c}{D1}\\
\hline
&$r < 18$ & 0.0392 & 0.0330 & 0.0632  & 0.0391& \\ 
&$18<r<19$& 0.0390 & 0.0430 & 0.0520  & 0.0533& \\ 
&$19<r<20$& 0.0391 & 0.0399 & 0.0366  & 0.0413&\\ 
&$20<r<21$& 0.0403 & 0.0471 & 0.0363  & 0.0665&\\ 
&$21<r<22$& 0.0652 & 0.0702 & 0.1051  & 0.1306&\\
\hline
&All & 0.0383 & 0.0338 & 0.0485 & 0.0307&
\enddata
\label{table_sigdist_ks}
\tablecomments{$\sigma_{\rm dist}$ and KS statistic results for CC2 and D1 ANN photo-z's for the validation set.}
\end{deluxetable}

\begin{figure*}
  \begin{center}
    \begin{minipage}[t]{81mm}
      \begin{center}
      \resizebox{81mm}{!}{\includegraphics[angle=0]{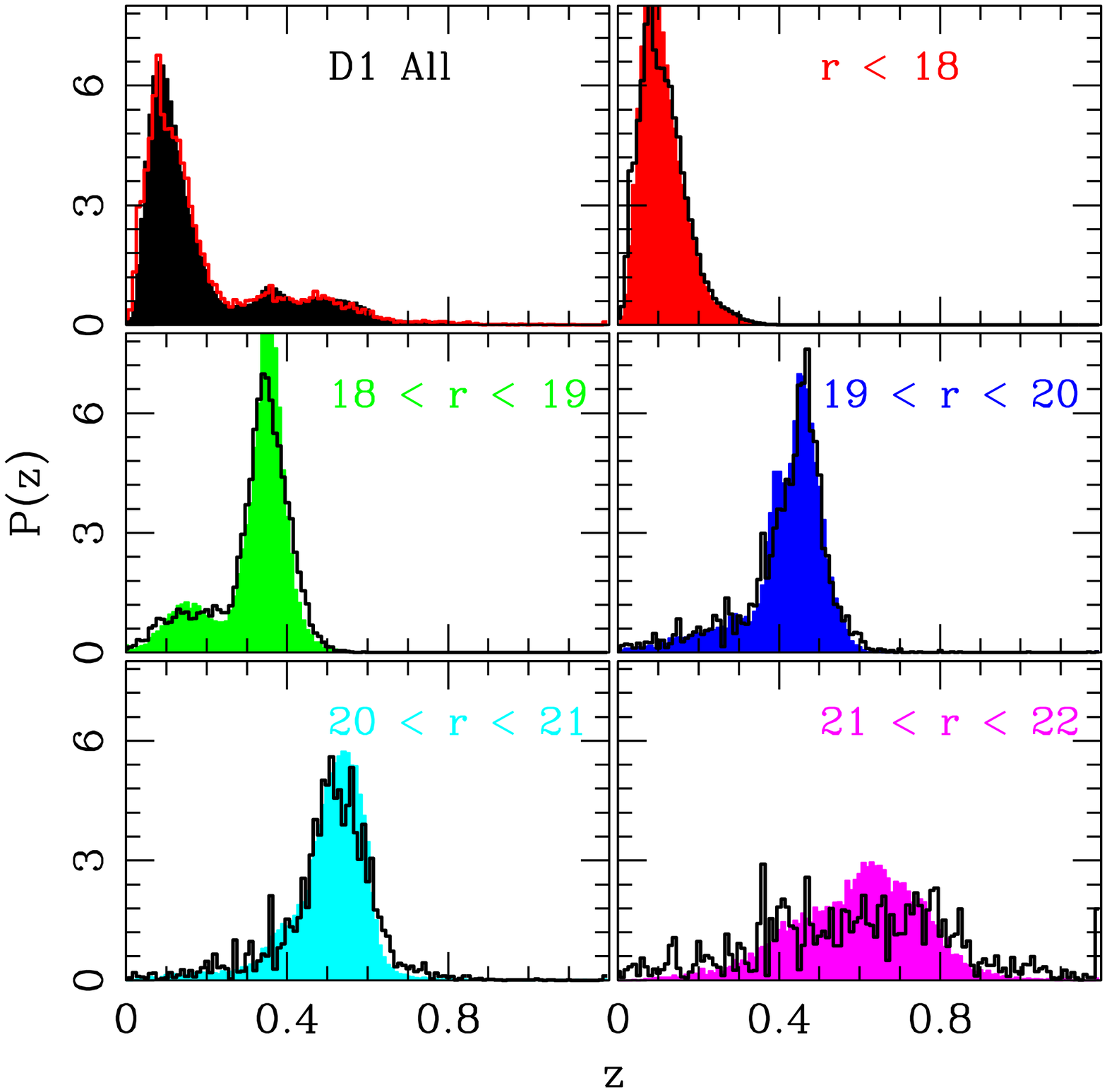}}
      \end{center}
    \end{minipage}
    \begin{minipage}[t]{81mm}
      \begin{center}
      \resizebox{81mm}{!}{\includegraphics[angle=0]{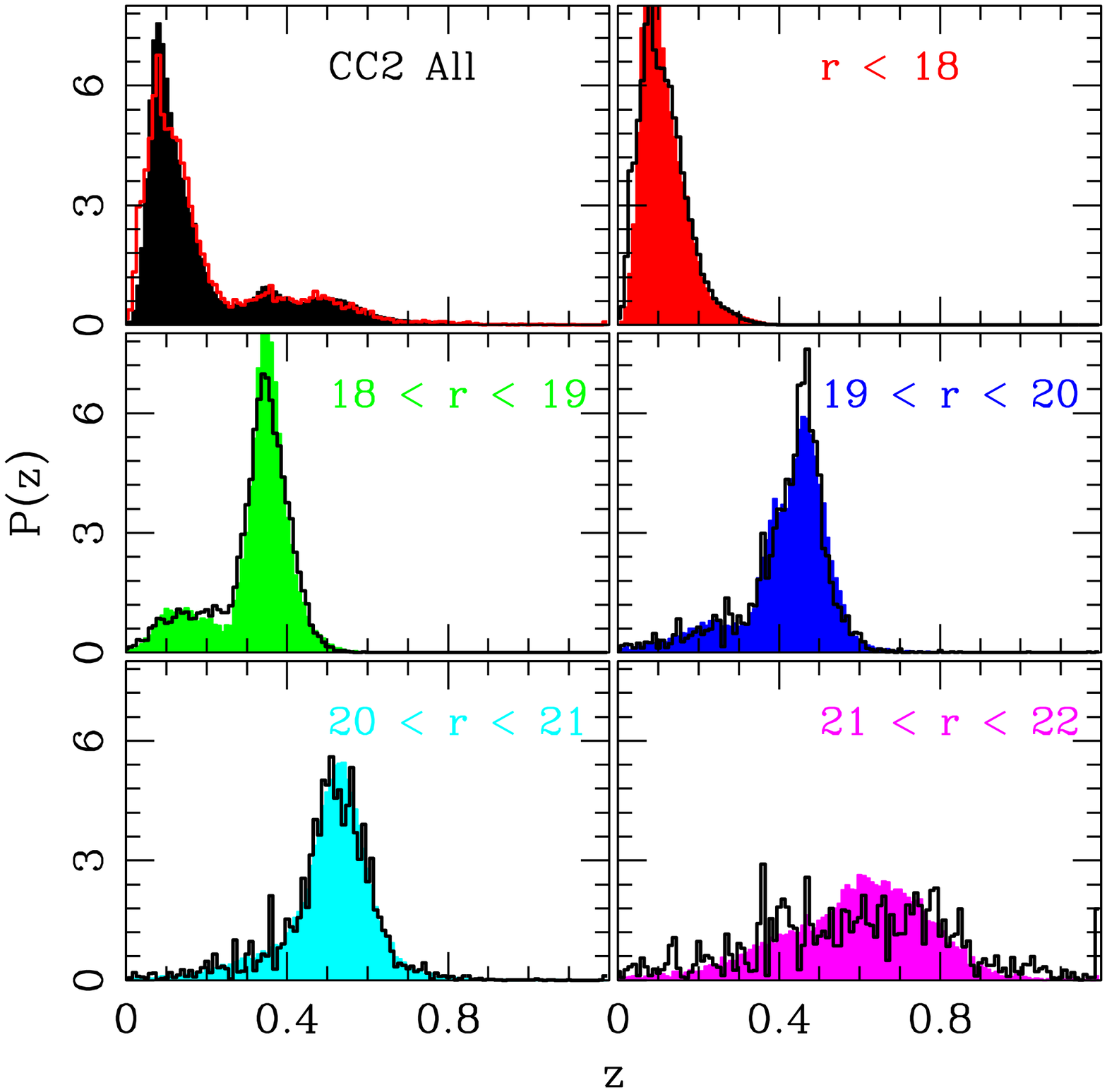}}
      \end{center}
    \end{minipage}
  \end{center}
  \caption{Redshift distributions for the galaxies in the 
    validation set for different $r$ magnitude bins. {\it Left panels:} ANN D1; 
    {\it right panels:} ANN CC2.
    The 
    colored regions indicate the ANN
    photo-z distributions, while the lines are
    the spectroscopic redshift distributions. By eye, 
    both ANN cases recover the true redshift distributions of the 
    validation set well, except
    in the faintest magnitude bin, where the photometric errors become large.
}\label{dndz.valid}
\end{figure*}

The redshift distributions for the validation set are shown in 
Fig.~\ref{dndz.valid} for the same bins of $r$ magnitude as in 
Table \ref{table_sigdist_ks}.
The D1 and CC2 \zphot \ distributions are shown 
in color,
and the solid curves correspond to the \zspec \ distributions.
The similarities between the \zphot \ and \zspec \ distributions 
are consistent with the results of 
Table \ref{table_sigdist_ks}.

In \S \ref{estdist}, we noted that the \zspec \ distribution of the 
spectroscopic sample, weighted to reproduce the color and magnitude 
distributions of the photometric sample, provides an estimate of the 
unknown redshift distribution of the photometric sample. The \zphot \ 
distribution for the photometric sample, computed using ANN D1 or CC2, provides 
another estimate of the true redshift distribution for the photometric 
sample, but one that we know suffers from bias (e.g., Fig. \ref{plot:statvsm}). 
While we have not shown that the weighted \zspec \ estimate of the 
redshift distribution is unbiased, it has the advantage that it makes 
direct use of the statistical properties of the photometric sample, and 
we believe it is our best estimate of the photometric sample redshift distribution.
Our final test of photo-z performance therefore compares the \zphot 
\ distribution for the photometric sample for the two ANN cases 
with the weighted \zspec \ distribution of the spectroscopic sample.
Agreement between the weighted \zspec \ distribution and either one of the 
\zphot \ distributions does not guarantee that they are correct, but 
it at least provides a useful consistency check.

\begin{figure*}
  \begin{center}
    \begin{minipage}[t]{81mm}
      \begin{center}
      \resizebox{81mm}{!}{\includegraphics[angle=0]{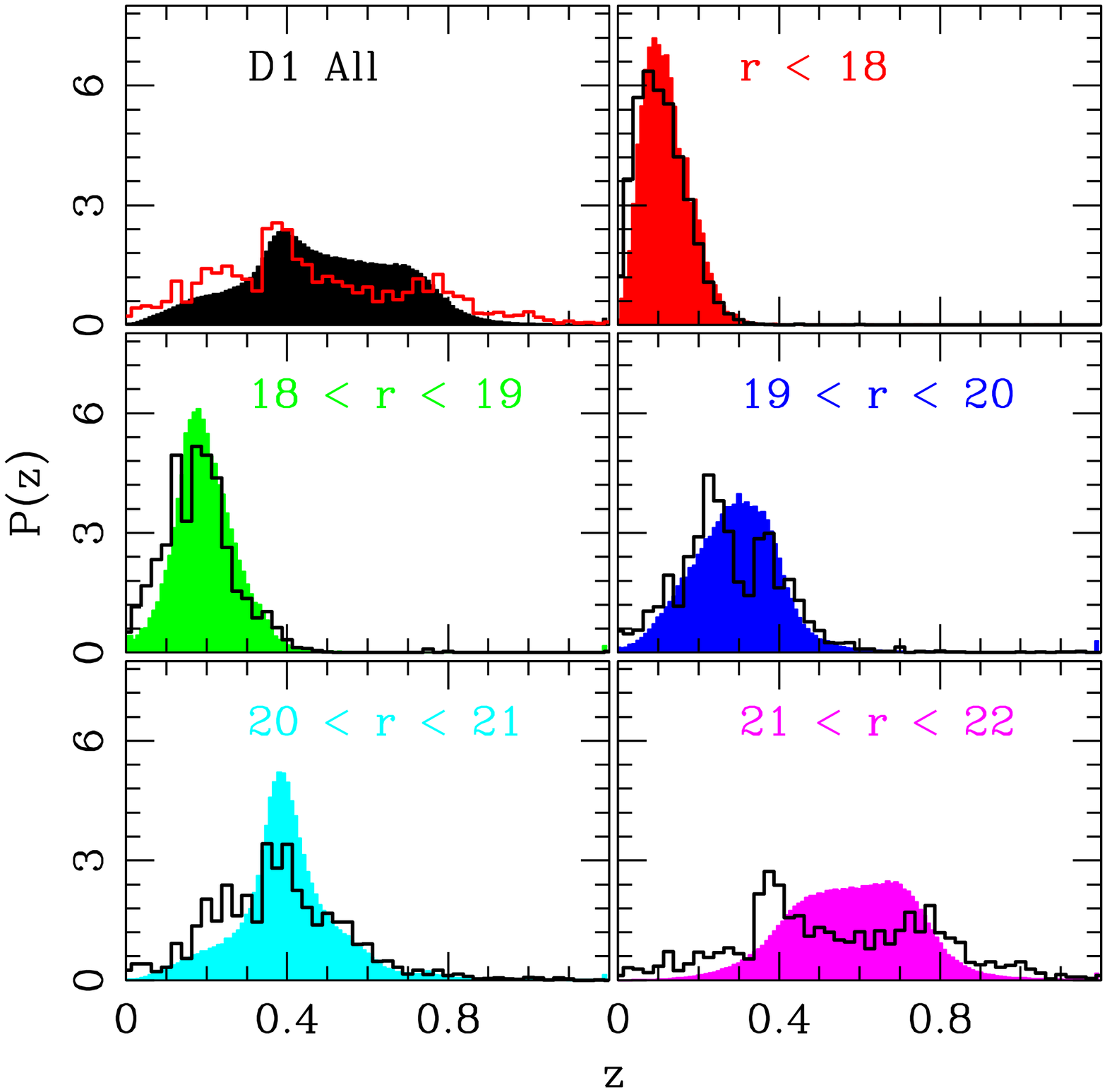}}
      \end{center}
    \end{minipage}
    \begin{minipage}[t]{81mm}
      \begin{center}
      \resizebox{81mm}{!}{\includegraphics[angle=0]{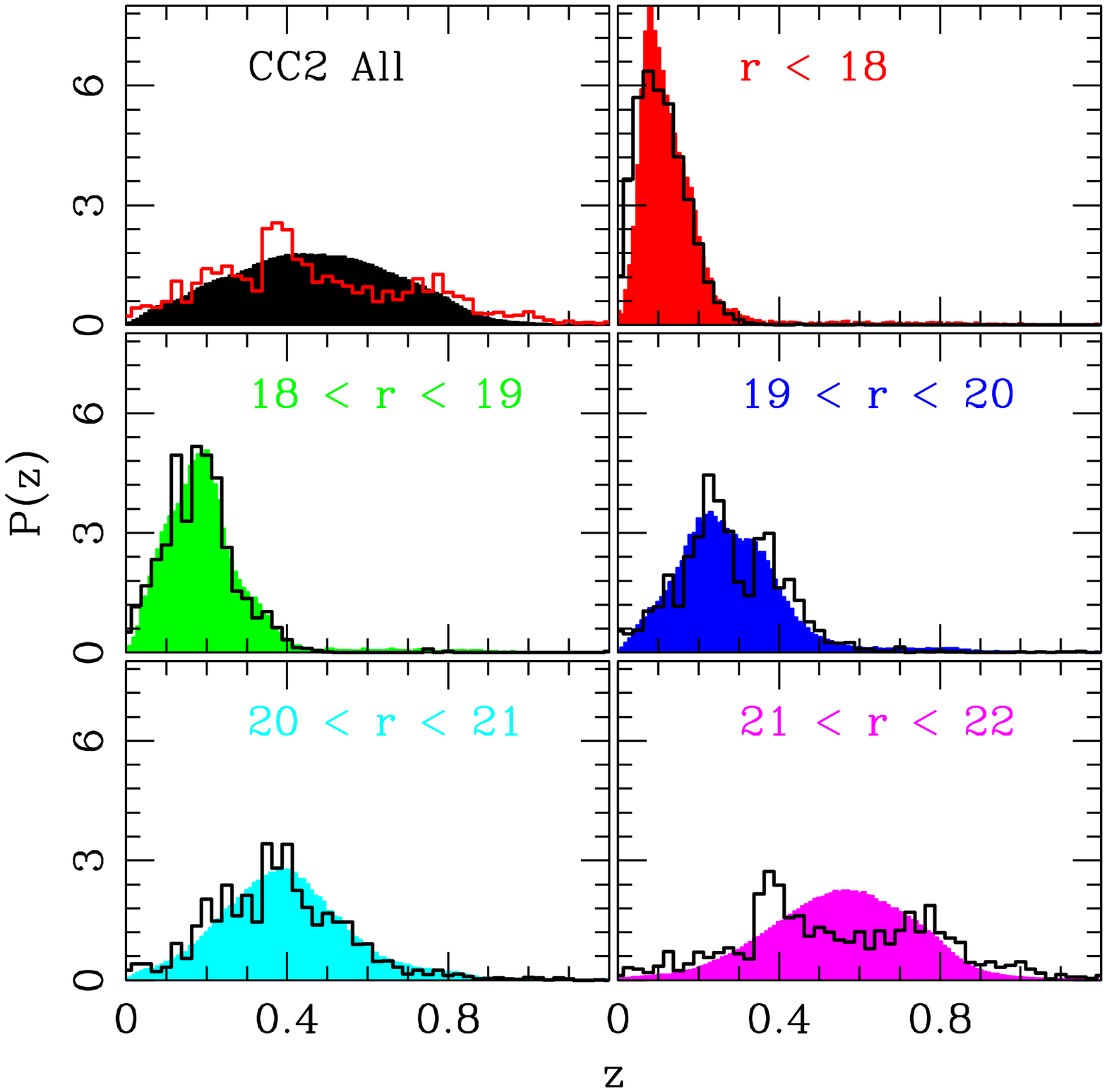}}
      \end{center}
    \end{minipage}
  \end{center}
  \caption{Estimated redshift distributions for a random subsample of 
    1\% of the galaxies in the 
    DR6 photometric sample in different $r$-magnitude bins. {\it Left panels:} 
    ANN D1; {\it right panels:} ANN CC2. Colors show the \zphot \ distributions.
    The lines show the estimated redshift distributions from the spectroscopic 
    sample weighted to match the magnitude and color distributions of the 
    photometric sample. 
    Even though the two ANN cases correctly recover the 
    validation set redshift distribution (Fig. \ref{dndz.valid}), 
    their photo-z 
    distributions for the photometric sample disagree. The photo-z distribution 
    for D1 shows a peak at 
    $z\sim0.4$ that results mainly from the $20 < r < 21$ bin.
    The CC2 distribution does not show such strong features, and in general it matches
    the weighted \zspec \ distribution better.
}\label{dndz.photo}
\end{figure*}

In Fig.~\ref{dndz.photo} we show the estimated redshift distributions of a
random subsample containing $\sim 1\%$ of the objects in the DR6 
photometric sample for both the CC2 and D1 ANN cases. 
The 
colored regions
correspond to the \zphot \ distributions, and the solid lines indicate 
the weighted \zspec \ distribution of the spectroscopic sample. 
The \zphot \ distributions for CC2 are closer matches to 
the weighted  \zspec \ distributions for $r>18$, and they do 
not show the peculiar features that the D1 photo-z distributions  
display, particularly at faint magnitudes. By the criterion of 
producing a more realistic redshift distribution for the photometric 
sample, the CC2 ANN estimator is preferred. 

\subsection{Photo-z Errors}

In order to test the quality of our photo-z error estimates
calculated with the NNE method, we introduce the concept of
empirical error. For a set of objects (within the validation set) with similar 
NNE error, 
$\sigma_{z}^{\rm NNE}$, the empirical error is defined as the $68\%$ 
width of the $|z_{\rm phot}-z_{\rm spec}|$ distribution for the set.
If the NNE estimator works properly, 
objects with similar NNE error should have similar underlying 
error distributions, i.e.,  
the NNE error should correlate
well with the empirical error.

\begin{figure*}
  \begin{center}
    \begin{minipage}[t]{81mm}
      \begin{center}
      \resizebox{81mm}{!}{\includegraphics[angle=0]{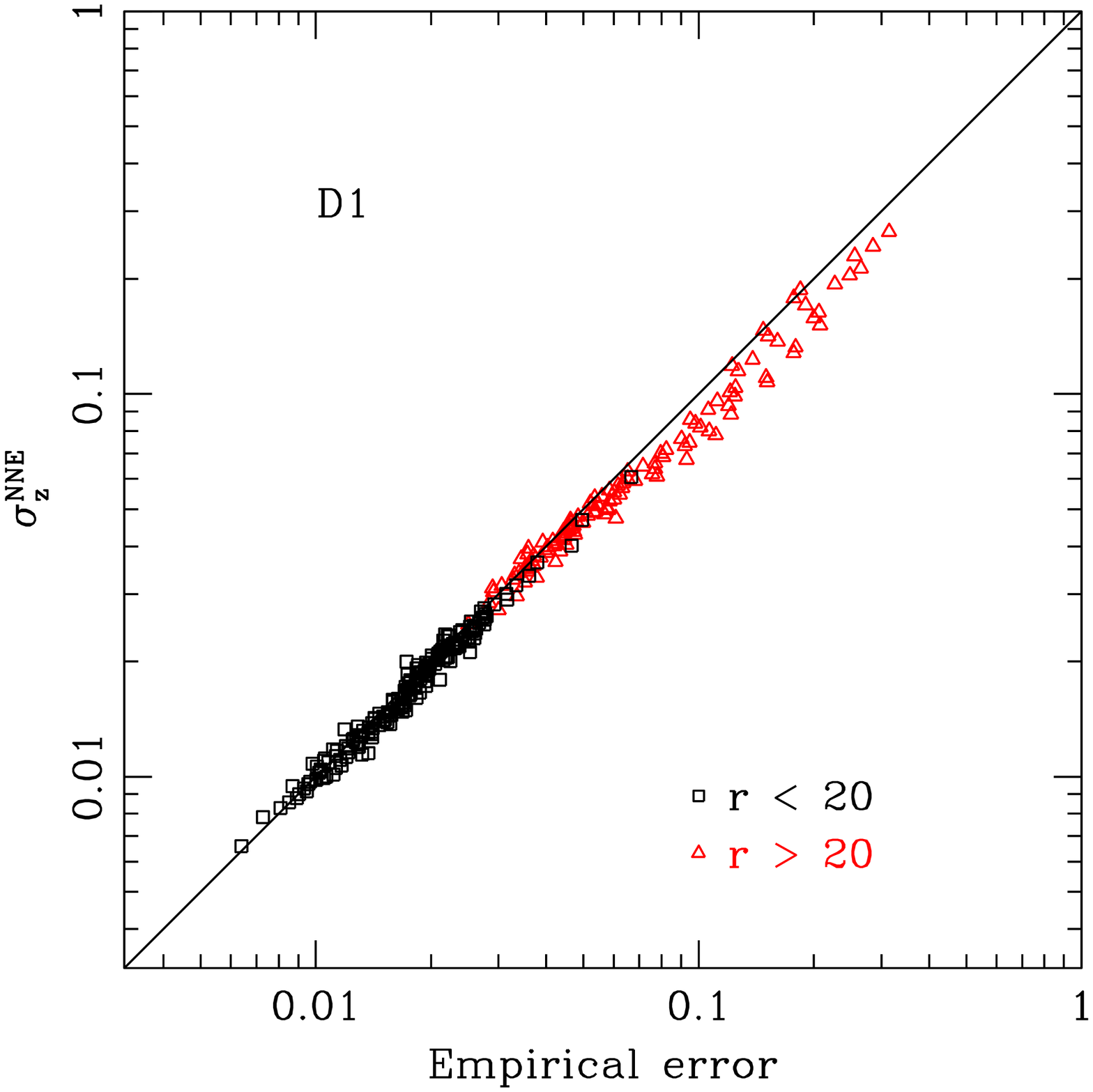}}
      \end{center}
    \end{minipage}
    \begin{minipage}[t]{81mm}
      \begin{center}
      \resizebox{81mm}{!}{\includegraphics[angle=0]{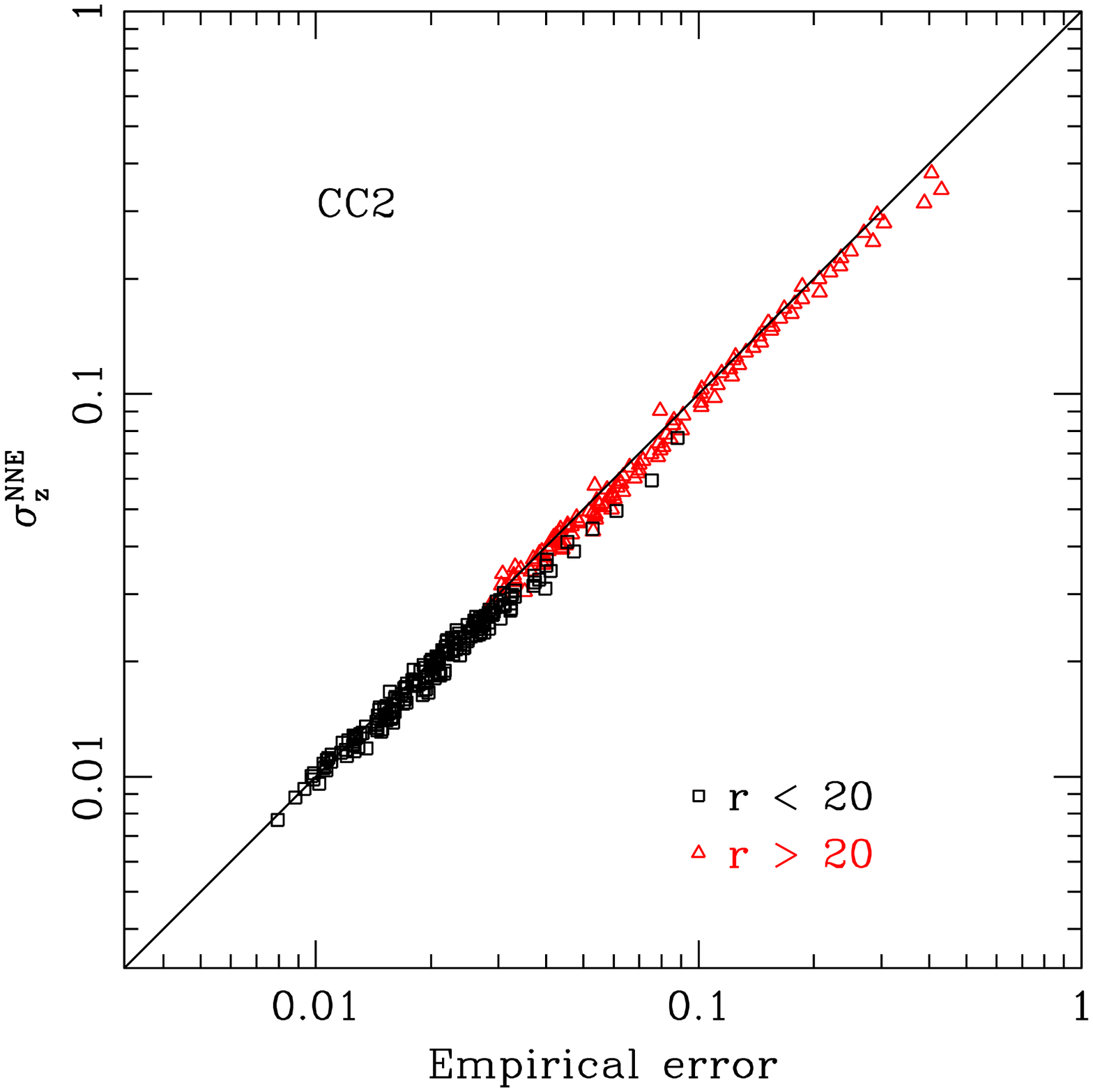}}
      \end{center}
    \end{minipage}
  \end{center}
 \caption{The estimated error from the NNE method, $\sigma_z^{\rm NNE}$, is 
   shown against the empirical error for objects in the validation set. 
   {\it Left panel:} D1 ANN; {\it right panel:} CC2 ANN. 
   Each point corresponds to a bin
   of $100$ objects with similar $\sigma_z^{\rm NNE}$.  
   The black squares show results for bright objects ($r < 20$),  
   the red triangles for faint objects ($r > 20$). As expected, faint 
   objects have larger errors, but 
   the NNE error correlates well with the empirical error over the full magnitude range.
 }\label{erer}
\end{figure*}

Fig.~\ref{erer} shows the performance of the photo-z error estimator
by plotting the computed NNE error $\sigma_{z}^{\rm NNE}$ as a function 
of the corresponding empirical error for the validation set. 
Results are shown for the D1 and CC2 ANN photo-z's.
The empirical error was calculated for bins containing $100$ objects 
with similar $\sigma_z^{\rm NNE}$.  
As expected, faint objects ($r > 20$) have larger errors than bright
objects ($r < 20$). 
The NNE estimated error correlates well with the 
empirical error even for the faint objects, indicating that the 
error estimator works properly for all magnitudes. 
The bulk of the bright objects have $\sigma_z^{\rm NNE}$ in the range
$0.01-0.04$, consistent with the overall {\it rms} photo-z scatter of 
$\sigma \sim 0.03$ indicated in Fig \ref{zpzs_valid_all}. 
Likewise, faint objects have $\sigma_z^{\rm NNE}$ in the range $0.02-0.3$, 
while $\sigma \sim 0.13$ for those objects. 
The NNE error is therefore a robust indicator of an object's
photo-z quality. In particular, we have carried out tests in which we 
cut objects with large NNE error from the sample and found that the 
remaining sample has smaller photo-z scatter and fewer catastrophic 
outliers. For applications in which 
photo-z precision is more important than 
completeness of the photometric sample, this can be a 
useful procedure. 

\begin{figure}
  \begin{center}
    \begin{minipage}[t]{81mm}
      \begin{center}
      \resizebox{81mm}{!}{\includegraphics[angle=0]{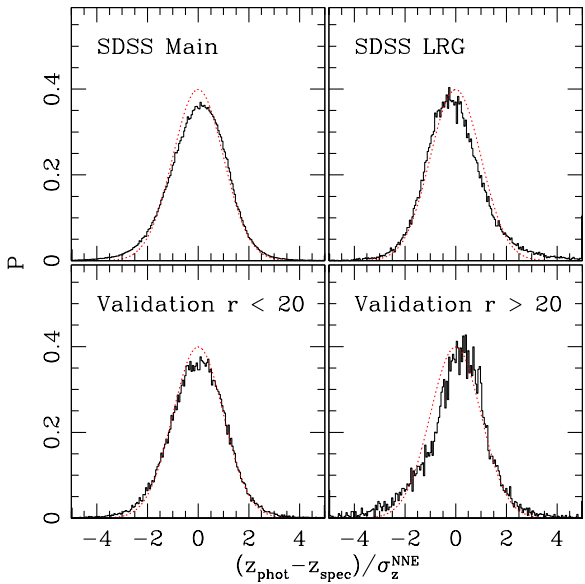}}
      \end{center}
    \end{minipage}
  \end{center}
 \caption{
   Distributions of 
   $(z_{\rm phot}-z_{\rm spec})/\sigma_{z}^{\rm NNE}$
   for objects in the spectroscopic sample, with photo-z's calculated
   using ANN D1; the   
   results for ANN CC2 are very similar. 
   The solid black lines are the data, and the dotted red lines are 
   Gaussians with zero mean and unit variance.  {\it Top left:} SDSS Main 
   spectroscopic sample; {\it top right:} SDSS LRG sample; {\it bottom 
     left:} validation-set galaxies with $r<20$; {\it bottom right:} validation-set
   galaxies with $r>20$. In all cases the photo-z errors 
   are reasonably well modeled by Gaussian distributions.
 }\label{gausser}
\end{figure}

In Fig.~\ref{gausser}, we plot the normalized error distribution, 
i.e., the distribution 
of $(z_{\rm phot}-z_{\rm spec})/\sigma_{z}^{\rm NNE}$, for objects 
in the spectroscopic sample, using the D1 ANN estimator. 
The solid black lines are the data, and the dotted red lines 
show Gaussian distributions with zero mean and unit variance. 
The upper panels show results for the galaxies in the SDSS Main 
and LRG spectroscopic samples. The lower panels show results for 
all validation-set galaxies, divided into bright 
($r < 20$) and faint ($r > 20$) samples.
These plots indicate that, averaged over the bulk of the spectroscopic 
sample, the photo-z estimates are nearly unbiased, the NNE error 
provides a good estimate of the true error, and the NNE error can be 
approximately interpreted as a Gaussian error in this average sense. 
Note that this does {\it not} imply that the photo-z error distributions in 
bins of magnitude or redshift are unbiased Gaussians: Figs. \ref{plot:statvsm} 
and \ref{plot:statvsz} show that they are not.

\section{Query Flags and Caveats} \label{rec}

When querying the SDSS data server to produce the photometric sample for 
which we estimated photo-z's, we set the most relevant flags needed to 
produce a clean galaxy sample.
However, some applications may require more stringent selection of objects. 
We advise users of the catalog to read the documentation about producing a clean 
galaxy sample on the SDSS 
website\footnote{ {\tt http://cas.sdss.org/dr6/en/help/docs/algorithm.asp} }.
In particular, users should consider requiring the BINNED1 (object detected at $> 5\sigma$) flag and removing 
objects with the NODEBLEND (object is a blend but deblending was not possible) flag. The various PHOTO flags 
are described in more details at the above  
website as well as in Appendix \ref{query}.

Finally, we note that the training of the photo-z estimators included only 
galaxies, not stars. As a result, photo-z estimates for 
stars that contaminate the photometric sample will be wrong, and cutting 
objects with low $z_{\rm phot}$ will not remove them. Our tests on 
star/galaxy separation in the photometric sample are briefly 
described in Appendix \ref{stargal}.

\section{Accessing the Catalog} \label{cat}

The photo-z catalog can be accessed from the 
{\tt photoz2} table in the DR6 context on the
SDSS CasJobs site, at {\tt http://casjobs.sdss.org/casjobs/}.
A query similar to the one in the Appendix provides all objects
for which we computed photo-z's. 
Alternatively, one can simply perform a query that searches for 
objects with a {\tt photoz2} entry.

In addition to the {\tt photoz2} table in the SDSS CAS, an independent 
{\tt photoz} table is also available, for which the photo-z's
have been computed using a template-based technique; see 
\cite{csa07, ade07}.

\section{Conclusions}\label{con}
We have presented a public catalog of photometric redshifts for the SDSS DR6 
photometric sample using  
two different photo-z estimates, CC2 and D1, based on the ANN method.
As a consistency check, we have also calculated photo-z's using the NNP method,
a nearest neighbor approach, which gives very good agreement with 
the ANN results. 
The CC2 and D1 photo-z results are comparable. For the validation set, the 
D1 photo-z estimates have lower photo-z scatter for bright galaxies ($r<20$), 
and scatter similar to but slightly smaller than that of 
CC2 for objects with $r>20$. Our tests indicate 
that the SDSS photo-z estimates are most reliable for galaxies 
with $r<20$   
and that the scatter increases significantly at fainter magnitudes. 
For faint galaxies ($r>20$), we recommend using the CC2 photo-z estimate, 
since the CC2 \zphot \ distribution most closely resembles the \zspec \ 
distribution for the validation set and the weighted \zspec \ estimate
for the redshift distribution of the photometric sample. 
For users who wish to use, for simplicity, a single photo-z estimator
over the full
magnitude range, we recommend using CC2.

Finally, we have demonstrated that the NNE error estimator, included in the 
public catalog, 
provides a reliable measure of the photo-z errors and that the overall scaled 
photo-z errors are nearly Gaussian. 

Funding for the DEEP2 survey has been provided by NSF grant AST-0071048 and AST-0071198. The data presented herein were obtained at the W.M. Keck Observatory, which is operated as a scientific partnership among the California Institute of Technology, the University of California and the National Aeronautics and Space Administration. The Observatory was made possible by the generous financial support of the W.M. Keck Foundation. The DEEP2 team and Keck Observatory acknowledge the very significant cultural role and reverence that the summit of Mauna Kea has always had within the indigenous Hawaiian community and appreciate the opportunity to conduct observations from this mountain.

Funding for the SDSS and SDSS-II has been provided by the Alfred P. Sloan Foundation, the Participating Institutions, the National Science Foundation, the U.S. Department of Energy, the National Aeronautics and Space Administration, the Japanese Monbukagakusho, the Max Planck Society, and the Higher Education Funding Council for England. The SDSS Web Site is {\tt http://www.sdss.org/}.

The SDSS is managed by the Astrophysical Research Consortium for the Participating Institutions. The Participating Institutions are the American Museum of Natural History, Astrophysical Institute Potsdam, University of Basel, University of Cambridge, Case Western Reserve University, University of Chicago, Drexel University, Fermilab, the Institute for Advanced Study, the Japan Participation Group, Johns Hopkins University, the Joint Institute for Nuclear Astrophysics, the Kavli Institute for Particle Astrophysics and Cosmology, the Korean Scientist Group, the Chinese Academy of Sciences (LAMOST), Los Alamos National Laboratory, the Max-Planck-Institute for Astronomy (MPIA), the Max-Planck-Institute for Astrophysics (MPA), New Mexico State University, Ohio State University, University of Pittsburgh, University of Portsmouth, Princeton University, the United States Naval Observatory, and the University of Washington. 

\appendix

\section{Data Query Code}\label{query}

Here we provide the SDSS database query used to obtain part of the catalog containing  
the photometric sample used in this paper. 
Notice that the query requires the TYPE flag to be set to 3 (galaxies) and 
selects objects with dereddened model magnitude  $r<22.0$ to reflect
the SDSS nominal detection limit. 
The query to obtain objects with Right Ascension (RA) in the 
range $[0,170)$ is

\vspace{0.8 cm}

{\tt
declare @BRIGHT bigint set @BRIGHT=dbo.fPhotoFlags('BRIGHT')

declare @SATURATED bigint set @SATURATED=dbo.fPhotoFlags('SATURATED')

declare @SATUR\_CENTER bigint set @SATUR\_CENTER=dbo.fPhotoFlags('SATUR\_CENTER') 
\vspace{0.5 cm}

declare @bad\_flags bigint set @bad\_flags=(@SATURATED|@SATUR\_CENTER|@BRIGHT)
\vspace{0.5 cm}

select 

objID, ra, dec,type,dered\_u,dered\_g,dered\_r,dered\_i,dered\_z,

petroR50\_u, petroR50\_g, petroR50\_r, petroR50\_i, petroR50\_z,

petroR90\_u, petroR90\_g, petroR90\_r, petroR90\_i, petroR90\_z

\vspace{0.5 cm}

into MyDb.all\_ra\_0\_170

FROM PhotoPrimary

WHERE ((flags \& @bad\_flags)) = 0 AND (dered\_r<=22.0) AND (ra>=0.0) AND (ra<170.0)

AND (type = 3)

}

\vspace{0.5cm}

Here we provide a brief description of the flags used in the query:
BRIGHT indicates that an object is a duplicate detection of an object with 
signal to noise greater  
than $200 \sigma$; SATURATED indicates that an 
object contains one or more saturated pixels; 
SATUR\_CENTER indicates that the object center is close to at least one 
saturated pixel.
Note that in selecting PRIMARY objects (using PhotoPrimary), 
we have implicitly selected objects 
that either do {\it not} have the BLENDED flag set 
or else have NODEBLEND set or nchild equal zero. 
In addition, the PRIMARY catalog contains no BRIGHT objects, so 
the cut on BRIGHT objects in the query above is in fact redundant. 
BLENDED objects have multiple peaks detected within them, which PHOTO  
attempts to deblend into several CHILD objects. 
NODEBLEND objects are BLENDED but no deblending was attempted on them, because
they are either too close to an EDGE, or too large, or one of 
their children overlaps an edge. A few percent of the objects in 
our photometric sample have NODEBLEND set; some users may wish to 
remove them.

We also suggest that users require objects to have the 
BINNED1 flag set. 
BINNED1 objects were detected at $\geq 5 \sigma$ significance 
in the original imaging frame.

The SDSS webpage\footnote{\tt{http://cas.sdss.org/dr5/en/help/docs/algorithm.asp?key=flags}} provides
further recommendations about flags, which we strongly recommend that users read.

\section{Tests on star-galaxy separation}\label{stargal}

We used the SDSS database TYPE flag to select the galaxy 
photometric sample for our photo-z catalogs. To study the robustness 
of the TYPE flag in separating galaxies from stars, we also 
carried out tests using an independent star-galaxy 
classifier. 
Here we briefly describe both of these techniques and show the results 
obtained on photometric and  spectroscopic samples. 

The TYPE flag is based on the star-galaxy separator in the SDSS PHOTO 
pipeline, 
described in \cite{lup01} and updated in \cite{aba04}. 
For a given object, the pipeline computes the PSF and cmodel 
magnitudes in each passband\footnote{http://www.sdss.org/dr5/algorithms/photometry.html}, 
where the cmodel magnitude is a measure of the flux using a 
composite of the best-fit de Vaucouleurs and exponential models of 
the light profile.  If the condition 

\begin{equation}
m_{PSF}-m_{cmodel} > 0.145 
\end{equation}

\noindent is satisfied, type is set to GALAXY for that band; 
otherwise, type is set to STAR.  The object's global TYPE is
determined by the same criterion, but now applied to the 
summed PSF and cmodel fluxes from all passbands in 
which the object is detected.
\cite{lup01} show that an earlier version of this simple 
cut works at the $95\%$ confidence level for SDSS objects brighter 
than $r=21$.

The second star-galaxy separator we tested is the galaxy probability
defined in \cite{scr02}.
The galaxy probability (hereafter $probgals$) is a Bayesian probability estimate that an object
is a galaxy (and not a star), given the object's magnitudes and 
concentration parameter. Here 
the concentration parameter is {\it not} 
the ratio of Petrosian radii but is 
defined as the difference between an
object's PSF and exponential-model $r$ magnitudes. 
This concentration parameter is close to zero for stars, is positive
for bright galaxies, and approaches zero as galaxies become fainter.  

We conducted some simple tests to compare these classification schemes.
If we set the Bayesian $probgals$ threshold to a value between 0.5 and 0.9, 
then both methods agree on the classification of 
more than $90\%$ of the objects for 
a random 1\% subset of the SDSS photometric sample. 
We also tested the methods on a spectroscopic sample of 29,229
galaxies and stars (counting independent photometric
measurements of each object) from the 2SLAQ and DEEP2 catalogs 
with $r < 22$.
Defining stars as objects with $z_{\rm spec}<0.01$, the sample 
contains 24,541 galaxies and 4,688 stars. We wish to compare 
this spectroscopic ``truth table'' with the photometric classification 
of the two methods and with a combined method that classifies 
an object as a galaxy if and only if both separators classify it as a 
galaxy. 
For the purposes of this test, we say that 
the Bayesian scheme classifies an object as a galaxy if 
$probgals>0.5$. We define galaxy 
completeness as the ratio of correctly identified
galaxies to the total number of galaxies in the spectroscopic 
sample.
Purity is defined as the ratio of correctly identified galaxies 
to the number of objects identified (correctly or not) as galaxies by the 
classifier. The purity depends in part on the relative numbers of 
galaxies and stars in the spectroscopic sample.

\begin{figure}
  \begin{center}
    \begin{minipage}[t]{81mm}
      \begin{center}
      \resizebox{81mm}{!}{\includegraphics[angle=0]{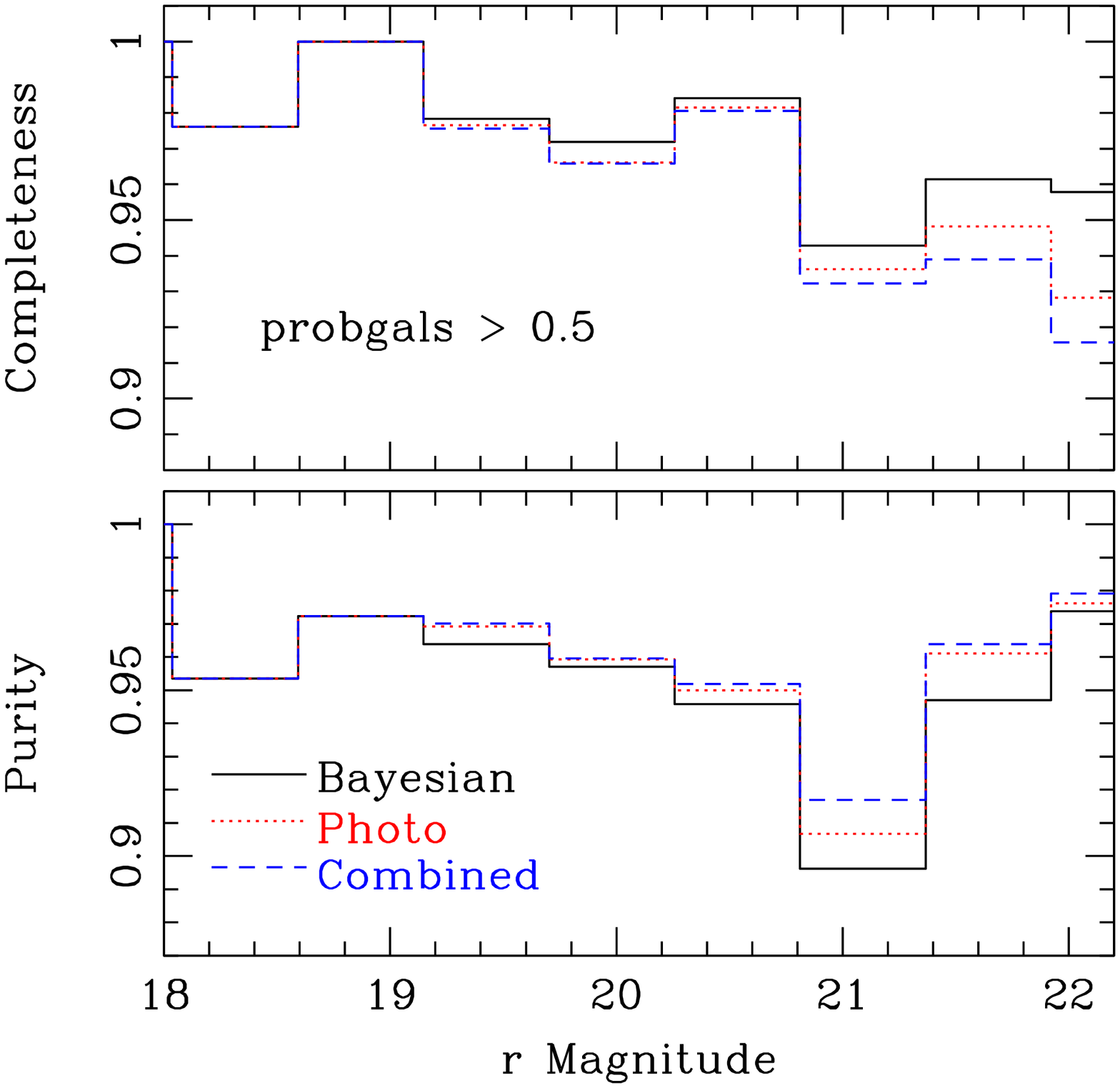}}
      \end{center}
    \end{minipage}
  \end{center}
 \caption{{\it Top panel:} completeness and {\it bottom panel:} purity 
for the 
Bayesian and PHOTO TYPE galaxy classifications as well as for a combination 
of the two, using a sample of galaxies with spectroscopic classification.
Results for the Bayesian separator have the $probgals$ lower bound set 
to $0.5$.}
\label{compur}
\end{figure}

Fig.~\ref{compur} shows the completeness and purity of the 
resulting galaxy catalogs in bins of $r$ 
magnitude for this spectroscopic sample.
Overall, the Bayesian separator and PHOTO TYPE 
produce similar results for galaxy purity and completeness. Moreover, 
the agreement between the two classification methods is quite good on 
an object-by-object basis. 
The 
Bayesian separator with {\it probgals} $\geq 0.5$ achieves slightly higher 
completeness and slightly lower purity. 
By varying the $probgals$ boundary, we could improve the purity of the 
Bayesian galaxy sample at the expense of degrading its completeness. 
We note that 
the best value of $probgals$ to use in defining a galaxy photometric 
sample depends on the scientific applications of the sample, i.e., 
on whether completeness or purity is the more important feature. 
In statistical applications, instead of defining a galaxy sample one 
can also choose to weight objects by 
their Bayesian probability \citep{scr02}. 

Based on this test, we conclude that 
the photometric sample for which we have estimated photo-z's has 
better than 90\% galaxy purity.

\section{Photometric Redshifts for SDSS DR5} 
\label{photdr5}

An earlier version of the photo-z catalog, produced for SDSS 
Data Release 5 (DR5), is publicly
available on the SDSS DR5 website (and is also called 
{\tt photoz2}).
The methods used to construct that photo-z catalog were similar to the
ones employed here for DR6, but the latter incorporates a number of 
important 
improvements. Here we briefly outline the differences between the two. 
We {\it strongly} recommend use of the DR6 photo-z catalog instead of the 
DR5 catalog. 

The photometric galaxy sample selection has improved from DR5 to DR6, 
because we used more stringent cuts in defining the DR6 sample.
The DR6 sample selection is described above in Appendix \ref{query}. 
The DR5 photometric galaxy sample selection required 
the cmodel and model $r$ magnitudes to lie in the ranges
$r_{\rm cmodel} \in (14.0,22.0)$ and 
$r_{\rm model} \in (13.5,22.5)$, and also required the value of   
the smear polarizability \citep{she04} to be $m_r>0.8$. Also, for DR5, 
star-galaxy separation used the Bayesian estimator (see 
Appendix \ref{stargal}) with the value $probgals >0.8$, while for DR6 
we used PHOTO TYPE. 
The additional cuts used for the DR6 catalog have produced 
a cleaner and more reliable galaxy sample.

\begin{deluxetable}{cccc}
\tablewidth{0pt}
\tablecaption{DR5 Catalog $flag$}
\startdata
\hline
\hline
\multicolumn{1}{c}{$flag$} 
& \multicolumn{1}{c}{N\textsuperscript{\b{o}} of Galaxies} 
& \multicolumn{1}{c}{Object Description}\\
\hline
- & $86.1$ million                 &All \\
0 & $12.6$ million               &\hspace{0.075 in}Complete \& bright\\ 
1 & $\hspace{0.06 in}0.6$ million &Incomplete \& bright\\ 
2 & $59.0$ million               &Complete \& faint \\ 
3 & $13.9$ million               &\hspace{-0.075 in}Incomplete \& faint \\
\enddata
\label{tableflags}
\tablecomments{The flag scheme for the DR5 catalog is based on object 
detection in some/all passbands and the $r$ magnitude. Incomplete objects are undetected in  
at least one of the passbands ($ugriz$) and faint objects have $r>20$.
}
\end{deluxetable}

\begin{deluxetable}{cccc}
\tablewidth{0pt}
\tablecaption{DR6 Catalog $flag$}
\startdata
\hline
\hline
\multicolumn{1}{c}{$flag$} 
& \multicolumn{1}{c}{N\textsuperscript{\b{o}} of Galaxies} 
& \multicolumn{1}{c}{Object Description}\\
\hline
- & $77.4$ million                 &All \\
0 & $11.5$ million               &bright\\ 
2 & $65.9$ million               &faint \\ 
\enddata
\label{tableflags6}
\tablecomments{The $flag$ scheme for the DR6 catalog is based solely on the 
on the $r$ magnitude: faint objects have $r>20$.
}
\end{deluxetable}

The DR5 photo-z catalog included 
a number of flags describing the expected photo-z
quality, shown in Table \ref{tableflags}. 
These flags were based on the detection or non-detection of the object in 
all passbands
and on the value of the $r$ model magnitude. An object was classified 
as bright (faint) if $r<20$ ($r>20$). An object was flagged as ``incomplete'' 
if it was not detected in all five SDSS passbands. Table \ref{tableflags} 
shows the corresponding flag values and the number of objects assigned 
each flag value. For the DR6 sample, given the stricter 
sample selection, a very small number of objects would have been 
classified as incomplete by the definition above, and they have 
been removed from the sample. As a result, for DR6, we only 
supply the bright/faint flag, as shown in Table \ref{tableflags6}.

The spectroscopic training set used for the DR6 photo-z catalog 
has important additions compared to
the one used for the DR5 catalog. In particular, 
for DR6 we added the DEEP2 spectroscopic catalog (which became 
publicly available), which made the training set more complete 
at faint magnitudes.
We also implemented more stringent spectroscopic quality cuts 
to the training set used for DR6.

Unlike the DR5 training set, the DR6 training set does not contain 
objects from the SDSS ``special'' plates, extra spectroscopic observations
designed to target specific objects for various scientific studies \citep{ade06}.
 In our tests, we find that
the lack of special plates does not result in any degradation of the 
photo-z quality.

The photo-z algorithm also changed from DR5 to DR6: we increased the 
number of hidden-layer nodes in the ANN and we added the concentration 
indices to the data inputs.
Our tests indicated that this leads to 
improved photo-z performance according to our metrics. 
In addition, the CC2 method differs from DR5 photo-z's further in that
CC2 uses only the color information and not the raw magnitudes.
For general purpose, full sample photo-z's, we recommend using CC2
photo-z's over both DR5 and D1 photo-z's.
Finally, 
we have carried out more extensive tests of the DR6 photo-z's than 
were done for DR5, increasing our confidence in the robustness of 
the photo-z estimates.

\bibliographystyle{apj}
\bibliography{ms}

\end{document}